\documentclass[aps,prd,groupedaddress,showpacs,nofootinbib,amssymb,balancelastpage,preprintnumbers,floatfix,10pt]{revtex4-2}

\usepackage{amsmath}
\usepackage{amssymb}
\usepackage{amsfonts}
\usepackage{bm}

\usepackage{graphicx}
\usepackage{xcolor}

\usepackage{dcolumn}

\usepackage{tikz}
\usetikzlibrary{matrix}

\allowdisplaybreaks[3]

\usepackage[colorlinks=true, linkcolor=blue!50!black, citecolor=blue!50!black, urlcolor=blue!60!black]{hyperref}

\begin{document}
	

\title{%
Purely quadratic non-Gaussianity from tachyonic instability: Primordial black holes and scalar-induced gravitational waves}

\author{He-Xu Zhang}\thanks{{\tt zhanghexu@ucas.ac.cn}}
\affiliation{School of Nuclear Science and Technology, University of Chinese Academy of Sciences, Beijing 100049, China}

\author{Mei Huang}\thanks{{\tt huangmei@ucas.ac.cn}}
\affiliation{School of Nuclear Science and Technology, University of Chinese Academy of Sciences, Beijing 100049, China}

\begin{abstract}
We investigate primordial black hole (PBH) formation in a cosmological scenario where curvature perturbations follow purely quadratic non-Gaussianity, $\zeta = A(\phi^2-\langle\phi^2\rangle)$, arising from tachyonic instability in multicomponent inflationary models. Within an extended Press-Schechter framework based on the compaction function, we derive the probability distribution of the linear compaction function and its asymptotic exponential tail, demonstrating that the PBH abundance is exponentially sensitive not only to the amplitude of perturbations but also to the correlation coefficient $\rho$ between the smoothed field and its radial gradient.
We further find that, in this tachyonic amplification scenario, the spectral width of the curvature power spectrum plays a decisive role in avoiding PBH overproduction: broad spectra yield mildly negative $\rho$ and fail to suppress PBH formation, while sufficiently narrow spectra drive $\rho \to -1$, resulting in exponential suppression while maintaining a sizable gravitational-wave signal.
Thermal inflation serves as a benchmark for asteroid-mass PBH dark matter and high-frequency scalar-induced gravitational waves potentially detectable by future space-based interferometers, but its typically broad spectra make it challenging to reconcile pulsar timing array observations with PBH constraints.
\end{abstract}
\maketitle
\vspace{-10pt}

\section{Introduction}

The recent pulsar timing array (PTA) data releases by the NANOGrav~\cite{NANOGrav:2023gor,NANOGrav:2023hde}, EPTA (in combination with InPTA) ~\cite{EPTA:2023fyk,EPTA:2023sfo,EPTA:2023xxk}, PPTA ~\cite{Reardon:2023gzh,Zic:2023gta,Reardon:2023zen}, and CPTA~\cite{Xu:2023wog} Collaborations have reported compelling evidence for the existence of a stochastic gravitational wave background (SGWB) in the nanohertz (nHz) band. These measurements exhibit a statistical preference for the Hellings-Downs correlation among pulsar pairs, widely regarded as the hallmark signature of an SGWB.
A leading cosmological interpretation of this SGWB is the stochastic background of scalar-induced gravitational waves (SIGWs) associated with large curvature fluctuations generated during inflation~\cite{Franciolini:2023pbf,Vagnozzi:2023lwo,Franciolini:2023wjm,Inomata:2023zup,Cai:2023dls,Wang:2023ost,Ebadi:2023xhq,Gouttenoire:2023nzr,Liu:2023ymk,Abe:2023yrw,Unal:2023srk,Yi:2023mbm,Firouzjahi:2023lzg,Salvio:2023ynn,You:2023rmn,Bari:2023rcw,Ye:2023xyr,HosseiniMansoori:2023mqh,Cheung:2023ihl,Balaji:2023ehk,Jin:2023wri,Bousder:2023ida,Das:2023nmm,Zhu:2023gmx,Ben-Dayan:2023lwd,Jiang:2023gfe,Liu:2023pau,Yi:2023tdk,Frosina:2023nxu,Bhaumik:2023wmw,Yuan:2023ofl,Gorji:2023sil,Liu:2023hpw,Chen:2024fir,Chen:2024twp}.
When the primordial curvature power spectrum experiences amplification on small scales, these scalar perturbations will source tensor modes via second-order nonlinear mode coupling upon horizon reentry during the radiation-dominated era, thereby generating an SGWB \cite{Tomita:1975kj,Matarrese:1993zf,Acquaviva:2002ud,Mollerach:2003nq,Ananda:2006af,Baumann:2007zm}.

Intriguingly, the large-amplitude scalar perturbations responsible for a detectable SIGW signal inevitably trigger gravitational collapse to form PBHs, provided that the local density fluctuations exceed a critical threshold upon horizon reentry. As a compelling nonbaryonic candidate for cold dark matter (DM), PBHs are nevertheless tightly constrained in their abundance over a wide range of masses through a variety of observational probes, including Hawking evaporation~\cite{Carr:2009jm,Clark:2016nst,Boudaud:2018hqb}, microlensing surveys~\cite{Macho:2000nvd,EROS-2:2006ryy,Griest:2013aaa,Mroz:2024mse,Mroz:2024wia,Mroz:2025xbl,Smyth:2019whb}, GW observations~\cite{LIGOScientific:2018mvr,LIGOScientific:2020ibl,KAGRA:2021vkt,LIGOScientific:2025slb}, dynamical effects~\cite{Monroy-Rodriguez:2014ula,Brandt:2016aco}, and accretion constraints from the CMB~\cite{Ali-Haimoud:2016mbv,Poulin:2017bwe,Serpico:2020ehh}.
Despite these constraints, a significantly unconstrained parameter space persists in the so-called \textit{asteroid-mass window}, spanning approximately $10^{-16}\,M_\odot \lesssim \text{M} \lesssim 10^{-10}\,M_\odot$. Within this mass range, PBHs can potentially account for the entire DM~\cite{Carr:2020xqk,Green:2020jor,Carr:2020gox}, and their corresponding SIGW signal naturally falls within the sensitivity band of space-based interferometers such as LISA~\cite{LISA:2017pwj,Babak:2021mhe}, Taiji~\cite{Luo:2019zal}, TianQin~\cite{Luo:2025sos,Luo:2025ewp}, DECIGO~\cite{Kawamura:2006up,Kawamura:2011zz}, and BBO~\cite{Crowder:2005nr,Corbin:2005ny}. Therefore, constructing a theoretical framework that explains the current PTA data and predicts GW signatures at higher frequencies, while generating a significant PBH DM population, is of considerable phenomenological interest (see, e.g., Refs.~\cite{Sasaki:2018dmp,Inomata:2018epa,DeLuca:2020agl,Vaskonen:2020lbd,Kohri:2020qqd,Wang:2019kaf,Abe:2022xur,Bhattacharya:2023ztw,Blas:2026xws,Cai:2018dig,Choudhury:2024kjj,Domenech:2021ztg,Escriva:2023uko,Ferrante:2022mui,Ferrante:2023bgz,Firouzjahi:2023xke,Frosina:2023nxu,Fu:2022ypp,Geller:2023shn,Inomata:2023drn,Inui:2024fgk,Iovino:2025cdy,Kitajima:2021fpq,Kumar:2025jfi,Pi:2021dft,Tada:2019amh,Shao:2022oqw,Zhang:2025kbu,Wang:2023ost,Wang:2024nmd,Balaji:2023ehk,Bhaumik:2025vlb,Cai:2025ksu,Choudhury:2023fwk,Choudhury:2024dzw,Dandoy:2023jot,Fei:2024vls,Figueroa:2023zhu,LISACosmologyWorkingGroup:2025vdz,Yang:2025vsq} for related works).

However, this interplay between the SIGW amplitude and the PBH abundance also introduces a severe theoretical tension. Interpreting the NANOGrav 15-year signal as SIGWs typically requires a large amplitude of the primordial power spectrum, $\mathcal{P}_\zeta \sim \mathcal{O}(10^{-2}-10^{-1})$ at the PTA scale $k\sim 10^6 \text{ Mpc}^{-1}$~\cite{Gorji:2023sil}.
Under the standard assumption of Gaussian primordial perturbations, the PBH DM fraction is exponentially sensitive to the power spectrum amplitude, $f_{\rm PBH} \propto \exp[-\mathcal{O}(1)/\mathcal{P}_\zeta]$. As a result, the large $\mathcal{P}_\zeta$ required to explain the PTA signal would generically yield an excessive PBH abundance, drastically violating existing observational bounds such as those from the LIGO-Virgo-KAGRA (LVK) merger rate limits~\cite{LIGOScientific:2018mvr,LIGOScientific:2020ibl,KAGRA:2021vkt,LIGOScientific:2025slb}.
Consequently, alleviating this overproduction tension naturally motivates considering non-Gaussianity (NG) in the primordial perturbations~\cite{Franciolini:2023pbf,Ellis:2023oxs}.

Indeed, incorporating these NGs is essential for computing the PBH abundance, since PBH formation is a rare-event phenomenon governed by the extreme tail of the probability distribution. In this tail region, even small deviations from Gaussian statistics can significantly modify the predicted abundance. These critical non-Gaussian corrections typically originate from two distinct sources. On the one hand, there are the intrinsic (or primordial) NGs present in the curvature perturbation $\zeta$ itself, which can be driven by early-Universe dynamics, such as in curvaton or multifield inflationary scenarios~\cite{Sasaki:2006kq,Enqvist:2009ww,Fonseca:2011aa,Pi:2021dft,Pi:2022ysn,Gow:2023zzp}. On the other hand, even in the absence of primordial NG in $\zeta$, the inherently nonlinear mapping between $\zeta$ and the density contrast $\delta$ at super-Hubble scales unavoidably induces NG~\cite{Harada:2015yda,DeLuca:2019qsy,Young:2019yug}.

Among the various scenarios generating primordial NG, a particularly appealing possibility arises in a class of models characterized by purely quadratic NG~\cite{Lyth:2011kj,Bugaev:2011qt,Lyth:2012yp,Dimopoulos:2019wew}. Specifically, the local curvature perturbation takes the form
\begin{equation}
\zeta(\mathbf{x}) = A \left( \phi^2(\mathbf{x}) - \sigma^2 \right)\,,
\label{eq:zeta_quadratic}
\end{equation}
where $\phi$ is a Gaussian random field with variance $\sigma^2 \equiv \langle\phi^2\rangle$, and $A$ is a model-dependent coefficient. Under this parametrization, the probability density function (PDF) $\mathbb{P}(\zeta)$ deviates substantially from Gaussian statistics and strictly follows a shifted $\chi^2$-distribution,
\begin{equation}
\mathbb{P}(\zeta) = \frac{1}{\sqrt{2\pi A \sigma^2 (\zeta + A\sigma^2)}} \exp\left( - \frac{\zeta + A\sigma^2}{2A\sigma^2} \right)\,.
\label{eq:PDF_zeta}
\end{equation}
The cosmological implications of Eq.~(\ref{eq:PDF_zeta}) depend critically on the sign of $A$. For $A > 0$, a scenario that might arise if $\zeta$ is generated after inflation by a curvaton-type mechanism~\cite{Gow:2023zzp}, the PDF exhibits a heavy exponential tail ($\mathbb{P}(\zeta) \sim \zeta^{-1/2} \exp[-\zeta/(2A\sigma^2)]$) that decays much slower than the standard Gaussian profile, thereby significantly promoting PBH formation. In contrast, for $A < 0$, since $\phi^2$ is strictly non-negative, the curvature perturbation is bounded from above, $\zeta \le \zeta_{\rm max} \equiv |A|\sigma^2$. This nontrivial feature effectively truncates the extreme positive tail of the distribution. Because PBHs form exclusively from rare regions with large positive overdensities, such a sharp cutoff can substantially suppress PBH production, thereby alleviating the aforementioned overproduction tension.

However, previous studies~\cite{Bugaev:2011wy,Lyth:2011kj,Bugaev:2011qt,Young:2013oia} have typically focused on the primordial NG encoded in the curvature perturbation $\zeta$, without fully accounting for additional NG generated by nonlinear gravitational dynamics. In addition, defining the collapse criterion in terms of $\zeta$ or the density contrast $\delta$ introduces intrinsic ambiguities. The absolute value of $\zeta$ has no local physical meaning, as it can be removed by a coordinate rescaling~\cite{Yoo:2020dkz}, while the use of $\delta$ requires a smoothing prescription, as simulations indicate that collapse is governed by the horizon-averaged density rather than peak value~\cite{Musco:2018rwt}. Instead, the compaction function, introduced in Ref.~\cite{Shibata:1999zs}, measures the mass excess within a spherical region and has been widely adopted as a robust criterion for collapse. In this work, we adopt a nonperturbative formalism~\cite{Gow:2022jfb,Gow:2023zzp} based on the compaction function, consistently incorporating both primordial NG and nonlinear gravitational contributions. The PBH abundance is then computed within an extended Press-Schechter framework by integrating the PDF $\mathbb{P}(\mathcal{C})$ above a critical threshold.

A natural dynamical origin for the purely quadratic NG in Eq.~(\ref{eq:zeta_quadratic}) arises from tachyonic instability in multicomponent inflationary settings, such as hybrid~\cite{Fonseca:2010nk,Gong:2010zf,Lyth:2011kj,Lyth:2012yp} or thermal inflation~\cite{Dimopoulos:2019wew,Lewicki:2021xku}. In these scenarios, a scalar field trapped at the origin in a false vacuum becomes tachyonic once the background system crosses a critical threshold. As the effective mass squared of the field turns negative, it triggers an explosive growth of long-wavelength fluctuations. During this tachyonic phase, the underlying symmetry of the potential forbids linear fluctuations, ensuring that the leading contribution arises entirely from quadratic terms. Consequently, the curvature perturbation naturally acquires the form in Eq.~(\ref{eq:zeta_quadratic}) with $A < 0$, which generally tends to suppress PBH formation. Importantly, this tachyonic amplification generates a sharply enhanced power spectrum on small scales, providing the seed fluctuations required for both SIGWs and PBHs in the scenarios considered here.

In this work, we investigate PBH formation and the associated SIGWs in scenarios where curvature perturbations are generated through tachyonic amplification, leading to purely quadratic NG. We first classify different classes of tachyonic amplification and show how the temporal structure of the instability determines the resulting spectral shape.

We then develop a nonperturbative framework based on the compaction function within an extended Press–Schechter approach, consistently incorporating both primordial NG and nonlinear gravitational effects. Within this framework, we derive the statistics of the compaction function and show that the PBH abundance exhibits a characteristic exponential sensitivity controlled not only by the amplitude of perturbations but also by the correlation coefficient $\rho$.

We further demonstrate that the spectral width of the curvature power spectrum governs the efficiency of PBH formation, and identify a regime in which PBH formation is exponentially suppressed while maintaining a sizable GW signal, thereby providing a robust mechanism to alleviate the PTA-–PBH tension. Finally, we present thermal inflation as a concrete realization and assess its implications for PBH DM and SIGW signals.

The rest of this paper is organized as follows. In Sec.~\ref{sec:Tachyonic_mechanism}, we present the generation of purely quadratic non-Gaussian curvature perturbations from tachyonic amplification and classify the resulting spectra. In Sec.~\ref{sec:PS_Formalism}, we formulate the Press–Schechter approach in terms of the compaction function for PBH formation. In Sec.~\ref{sec:SIGW}, we compute the associated SIGWs and confront the predictions with PTA observations. In Sec.~\ref{sec:thermal_inflation}, we present a concrete realization based on thermal inflation. Finally, Sec.~\ref{sec:summary} summarizes our conclusions.

\section{Tachyonic Amplification and Purely Quadratic Non-Gaussian Perturbations}
\label{sec:Tachyonic_mechanism}

A broad class of early-Universe scenarios generates scalar perturbations through a temporary tachyonic instability, which occurs when the effective mass-squared of a scalar field becomes negative for a finite period. Representative examples include waterfall transitions in hybrid inflation, thermally triggered symmetry breaking, or transient tachyonic phases of spectator fields. Despite their distinct underlying realizations, these mechanisms share a common dynamical feature and can therefore be described within a unified framework.

To illustrate this, we consider a canonical scalar field $\phi$ (e.g., the waterfall field in hybrid inflation~\cite{Fonseca:2010nk,Gong:2010zf,Lyth:2011kj,Lyth:2012yp}) evolving in a Friedmann-Lema\^{i}tre-Robertson-Walker (FLRW) background. The linear Fourier mode $\phi_k(t)$ obeys the equation of motion\footnote{Strictly speaking, $\phi_k$ denotes the Fourier mode of the fluctuation $\delta\phi$ around the homogeneous background $\bar{\phi}$. In the scenarios considered here the field is stabilized at the origin prior to the tachyonic instability, i.e. $\bar{\phi}\simeq0$, so we simply denote the fluctuation mode by $\phi_k$.}
\begin{equation}
    \ddot{\phi}_k + 3H\dot{\phi}_k + \left(\frac{k^2}{a^2} + m_{\rm eff}^2(t)\right)\phi_k = 0\,,
\end{equation}
where the overdot denotes the derivative with respect to the cosmic time $t$.

In a quasi-de Sitter background, it is convenient to remove the Hubble friction term by introducing the rescaled variable $u_k \equiv a^{3/2}\phi_k$, for which the equation of motion becomes
\begin{equation}
\ddot u_k + \Omega_k^2(t) u_k = 0\,,\quad \text{with} \quad \Omega_k^2(t) \equiv \frac{k^2}{a^2} + m_{\rm eff}^2(t) - \frac{9}{4}H^2.
\end{equation}
A tachyonic regime is signaled by a negative effective mass-squared, $m_{\rm eff}^2(t) < 0$, while exponential growth occurs for modes satisfying $\Omega_k^2 < 0$.
Defining the effective tachyonic scale
\begin{equation}
M_{\rm eff}(t) \equiv \sqrt{\frac{9}{4}H^2 + |m_{\rm eff}^2(t)|}\,,
\end{equation}
the squared growth rate can be written as
\begin{equation}
\mu_k^2(t) \equiv -\Omega_k^2(t) = M_{\rm eff}^2(t) - \frac{k^2}{a^2(t)}.
\end{equation}
Accordingly, the instability band is approximately given by $k < a(t) M_{\rm eff}(t)$, within which long-wavelength modes undergo exponential amplification.

The adiabatic growth regime is characterized by the WKB condition $|\dot{\Omega}_k / \Omega_k^2| \ll 1$, which can be expressed parametrically as
\begin{equation}
\frac{H}{M_{\rm eff}(t)}\left(\frac{k}{a(t)M_{\rm eff}(t)}\right)^2
\ll \left[1-\left(\frac{k}{a(t)M_{\rm eff}(t)}\right)^2\right]^{3/2}, \qquad
\frac{1}{M^2_{\rm eff}(t)}\frac{\mathrm{d}M_{\rm eff}(t)}{\mathrm{d}t}
\ll \left[1-\left(\frac{k}{a(t)M_{\rm eff}(t)}\right)^2\right]^{3/2}.
\end{equation}
Under these conditions, the growing mode admits a WKB-type solution,
\begin{equation}
\phi_k(t) \simeq \frac{C_k}{\sqrt{2a^3(t)\mu_k(t)}}
\exp\left( \int_{t_{\rm in}}^t \mu_k(t') \,\mathrm{d}t' \right),
\end{equation}
where $t_{\rm in}$ denotes the onset of the adiabatic growth era. The dimensionless power spectrum, $\mathcal{P}_\phi(k) \propto k^3 |\phi_k|^2$, is determined by the time at which the tachyonic amplification terminates, denoted by $t_{\rm end}$.

In this work, we classify tachyonic amplification mechanisms into two classes depending on the temporal structure of the growth rate $\mu_k(t)$.

(i) \textit{Monotonic amplification.}
In this regime, the tachyonic mass-squared approaches a constant negative value $-m_0^2$ at late times, resulting in a monotonically increasing growth rate.
Once a mode enters the tachyonic band, the subsequent quasi-de Sitter expansion rapidly redshifts the gradient term, so that $k^2/a^2 \ll M_{\rm eff}^2(t)$. Consequently, the growth rate quickly becomes effectively independent of $k$, approaching $\mu_k(t) \simeq \sqrt{9H^2/4+m_0^2}\equiv \nu_0 H$.
For sufficiently small $k$, modes enter the tachyonic band very early, but the growth rate is still small at that time. Most of the amplification is accumulated during a later period when $\mu_k(t)$ has reached its asymptotic value, which is common to all modes. As a result, the amplification exponent becomes approximately independent of $k$, leading to the universal infrared scaling $\mathcal{P}_\phi(k)\propto k^3$. For larger $k$, modes enter the tachyonic band later, when the growth rate is already sizable. In this regime modes enter the tachyonic band sequentially according to $k/a(t_k) \simeq \nu_0 H$, yielding a $k$-dependent onset time $t_{\rm in}(k) \simeq H^{-1} \ln k + \text{const.}$.
The resulting amplification factor can then be approximated as
\begin{equation}
\mathcal{A}_k \sim \exp\left(\int_{t_{\rm in}(k)}^{t_{\rm end}} \nu_0 H \mathrm{d}t\right) \propto \exp\left(-\nu_0\ln k\right) = k^{-\nu_0} \,.
\end{equation}
Incorporating the phase-space volume factor, the power spectrum in this tachyonic regime naturally develops a power-law UV tail:
\begin{equation}
    \mathcal{P}_\phi(k) \propto k^3 |\mathcal{A}_k|^2 \sim k^{3 - 2\nu_0}.
\end{equation}
For numerical analysis, it is convenient to adopt a smooth broken power-law (BPL) parametrization,
\begin{equation}
    \mathcal P^{\rm BPL}_\phi(k) = \mathcal A_\phi\,\frac{(3+\beta)^s}{\left[\beta\left(k/k_\ast\right)^{-3/s}+3\left(k/k_\ast\right)^{\beta/s}\right]^s}\,,
    \label{eq:BPL_spectrum}
\end{equation}
which reproduces the asymptotic scalings $\mathcal P_\phi\propto k^3$ for $k\ll k_\ast$
and $\mathcal P_\phi\propto k^{-\beta}$ for $k\gg k_\ast$.
Here the UV slope $\beta$ is determined by the underlying tachyonic mass scale (e.g. by the ratio $m_0/H$), while $s$ determines the smoothness of the transition.

(ii) \textit{Nonmonotonic amplification.}
In this class, the growth rate increases, reaches a local maximum, and subsequently decreases, due to backreaction effects or time-dependent couplings to a rolling background field~\cite{Dimastrogiovanni:2016fuu,Kumar:2025jfi}.
During the tachyonic phase, the gradient term rapidly redshifts, so that $\mu_k(t) \simeq M_{\rm eff}(t)$.
To be concrete, we parametrize this mass around its maximum at $t_{\rm max}$ as
\begin{equation}
    M_{\rm eff}(t) \simeq \nu_{\rm max} H \left[ 1 - \frac{(t - t_{\rm max})^2}{2\sigma_t^2} \right] \,, \quad \nu_{\rm max} \equiv \sqrt{\frac{9}{4}+\frac{m_{\rm max}^2}{H^2}}\,.
\end{equation}
where $m_{\rm max}$ denotes the maximal tachyonic mass scale and $\sigma_t$ characterizes the duration of the instability.
The power spectrum is determined by the total spectral exponent, which combines the phase-space volume and the dynamical WKB amplification:
\begin{equation}
\ln \mathcal{P}_\phi(k) \simeq 3 \ln k + 2 \int_{t_{\text{in}}(k)}^{t_{\text{end}}} M_{\rm eff}(t)\,\mathrm{d}t\,.
\end{equation}
Using $\mathrm{d}t_{\text{in}}/\mathrm{d}\ln k \simeq H^{-1}$, the location of the spectral peak $k_\ast$ is determined by the stationary condition
\begin{equation}
\left.\frac{\mathrm{d} \ln \mathcal{P}_\phi}{\mathrm{d} \ln k}\right|_{k=k_\ast} \simeq 0 \quad \Rightarrow \quad M_{\rm eff}(t_\ast) \simeq \frac{3}{2}H\,,\quad \text{with} \quad t_\ast \equiv t_{\rm in}(k_\ast)\,.
\end{equation}
The curvature of the spectrum at the peak is then given by
\begin{equation}
\left. \frac{\mathrm{d}^2\ln\mathcal{P}_\phi}{\mathrm{d}(\ln k)^2} \right|_{k=k_\ast} \simeq -\frac{2}{H^2} \left. \frac{\mathrm{d}M_{\rm eff}(t)}{\mathrm{d}t} \right|_{t=t_\ast} = \frac{2\nu_{\rm max}}{H\sigma_t^2}(t_\ast - t_{\rm max}) \equiv -\frac{1}{\Delta^2} < 0\,.
\end{equation}
A saddle-point approximation around $k_\ast$ directly yields a log-normal (LN) spectrum. For practical applications, it is convenient to express this in the normalized form
\begin{equation}
    \mathcal{P}^{\rm LN}_\phi(k) = \frac{\mathcal{A}_{\phi}}{\sqrt{2\pi}\Delta} \exp\!\left[ -\frac{(\ln k - \ln k_\ast)^2}{2\Delta^2} \right],
    \label{eq:LN_spectrum}
\end{equation}
where $k_\ast$ and $\mathcal{A}_\phi$ denote the peak scale and amplitude of the source-field power spectrum, respectively, while $\Delta$ controls its width.

Having specified the power spectrum of the source field $\phi$, we now describe how these fluctuations are converted into curvature perturbations. In the scenario considered here, the leading contribution to the curvature perturbation is generated by the quadratic fluctuation of the tachyonically amplified field. We therefore parametrize the curvature perturbation as
\begin{equation}
    \zeta(\mathbf{x}) = A(\phi^2(\mathbf{x}) - \langle\phi^2\rangle)\,,
    \label{eq:purely_NG}
\end{equation}
where $A$ is a model-dependent coefficient to be determined.

This relation can also be understood from the perspective of the $\delta N$ formalism\footnote{In the presence of a tachyonic instability, the classical expansion history $N(\phi)$ may become singular near the symmetric point $\bar{\phi}=0$. As illustrated by the waterfall transition in hybrid inflation~\cite{Fonseca:2010nk}, this singularity can be regularized by separating the field into long- and short-wavelength components, $\phi_L$ and $\phi_S$, and applying the $\delta N$ expansion to the short-mode averaged quantity $\langle N\rangle_L$.}. In this framework, the curvature perturbation is given by $\zeta\equiv\delta N$ and can be expanded as
\begin{equation}
    \zeta \simeq N_{,\phi}\delta\phi + \frac12 N_{,\phi\phi}\delta\phi^2 + \cdots \,.
\end{equation}
Since the tachyonic dynamics is invariant under the transformation $\phi \rightarrow -\phi$, the expansion history is an even function of the source field around the symmetric point, implying
$\left.N_{,\phi}\right|_{\bar{\phi}=0} = 0\,.$
Therefore, the leading contribution to the curvature perturbation arises at second order in the source-field fluctuation, providing a natural realization of purely quadratic NG. This interpretation is consistent with the direct derivation based on the energy-density perturbation in the spatially flat gauge adopted in this work.

The resulting curvature power spectrum is obtained from the convolution in momentum space~\cite{Lyth:1991ub},
\begin{equation}
    \mathcal{P}_\zeta(k) = \frac{A^2k^3}{2\pi} \int \mathrm{d}^3q \frac{\mathcal{P}_\phi(q) \mathcal{P}_\phi(|\mathbf{k}-\mathbf{q}|)}{q^3 |\mathbf{k}-\mathbf{q}|^3} \,.
\end{equation}

\section{Extended Press-Schechter Formalism and PBH Formation}
\label{sec:PS_Formalism}

In this section we compute the PBH abundance in the presence of purely quadratic NG within the extended Press-Schechter framework. 
We first introduce the compaction function as the collapse criterion, derive its statistical distribution generated by quadratic curvature perturbations in Eq.~(\ref{eq:purely_NG}), and evaluate the resulting PBH abundance.

\subsection{Compaction function}

PBHs are expected to form from rare, large-amplitude peaks in the primordial density field. For Gaussian statistics, peak theory implies that such high peaks~\cite{Bardeen:1985tr,Escriva:2021aeh} are approximately spherical, motivating the use of the spherical collapse approximation. In the present work, we assume that this remains a reasonable first-order description even in the strongly non-Gaussian regime considered here, following Ref.~\cite{Gow:2022jfb}. We therefore neglect angular dependence and restrict our analysis to the spherically symmetric monopole ($\ell=0$) of the curvature perturbation.

We consider the superhorizon evolution of curvature perturbations on a spatially flat FLRW background. In the gradient expansion approximation, the perturbed metric at leading order in the uniform-energy density gauge takes the form,
\begin{equation}
    \mathrm{d}s^2 = - \mathrm{d}t^2 + a^2(t) e^{2\zeta(r)} \left(\mathrm{d}r^2 + r^2\mathrm{d}\Omega^2\right),
\end{equation}
where $a(t)$ is the scale factor and $\zeta(r)$ is the curvature perturbation.

As discussed in the introduction, the curvature perturbation or the local density contrast does not always provide a reliable criterion for PBH formation in this regime. We therefore employ the compaction function $\mathcal{C}(r,t)$, a more robust measure defined by the mass excess enclosed within a sphere of areal radius $R(r,t)$ relative to the homogeneous background value~\cite{Shibata:1999zs},
\begin{equation}
	\mathcal{C}(r,t) \equiv 2G \frac{\left[ M_{\text{MS}}(r, t) - M_b(r, t) \right]}{R(r, t)} \,,
\end{equation}
where $M_\mathrm{MS} = \int_0^R\,4\pi\rho R'^2 \mathrm{d}R'$ is the Misner-Sharp mass and $M_b = \frac{4\pi}{3}\rho_b R^3$ is its background counterpart. The corresponding areal radius is
\begin{equation}
    R(r, t) = a(t) r e^{\zeta(r)} \,.
\end{equation}

On superhorizon scales, the density contrast is related to the curvature perturbation by the nonlinear relation~\cite{Harada:2015yda,Kawasaki:2019mbl,Young:2019yug,DeLuca:2019qsy},
\begin{equation}
\label{eq:density_contrast}
    \delta(r,t) \equiv \frac{\delta\rho}{\rho_b} = - \frac{4(1+w)}{5+3w}\frac{1}{a^2H^2}e^{-5\zeta(r)/2}\nabla^2 e^{\zeta(r)/2}\,.
\end{equation}
Substituting Eq.~\eqref{eq:density_contrast} into the definition of the compaction function, one finds that during radiation domination ($w=1/3$) the compaction function reduces to~\cite{Musco:2018rwt,Young:2019yug}
\begin{equation}
\label{eq:non-linear-Com}
	\mathcal{C}(r) = \frac{2}{3}\left[1-\left(1+r \zeta'(r)\right)^2\right] = \mathcal{C}_\ell(r) -\frac{3}{8}\mathcal{C}_\ell(r)^2\,,
\end{equation}
where the linear component,
\begin{equation}
\mathcal{C}_\ell(r) = -\frac{4}{3}\,r\,\zeta'(r)
\end{equation}
is obtained from the leading-order approximation of Eq.~\eqref{eq:density_contrast}. Notably, the time dependence cancels out in the final expression, rendering $\mathcal{C}(r)$ conserved on superhorizon scales, in contrast to the evolving density contrast.

From Eq.~\eqref{eq:non-linear-Com}, the compaction function reaches its maximum value $\mathcal{C}_{\rm max}=2/3$ or $\mathcal{C}_{\ell,{\rm max}}=4/3$, which marks the boundary between Type I and Type II perturbations~\cite{Kopp:2010sh}. Type I perturbations correspond to $\mathcal{C}_\ell\le4/3$, for which the areal radius increases monotonically with the comoving radius, whereas Type II perturbations correspond to $\mathcal{C}_\ell>4/3$, leading to a nonmonotonic areal radius associated with a “separate-universe” configuration.
We specifically exclude Type II perturbations from our calculations because the probability of such extreme overdensities are generally exponentially suppressed relative to Type I perturbations, and their gravitational collapse mechanisms are currently not well understood~\cite{Kopp:2010sh}.

\subsection{Probability distribution of purely quadratic perturbations}
Under the spherical symmetry assumption, the radial profile of the underlying Gaussian field $\phi(r)$, which sources the curvature perturbation $\zeta$, is given by
\begin{equation}
\phi(r) \equiv \phi_{00}(r) = \int \frac{\mathrm{d}^3\mathbf{k}}{(2\pi)^3} j_0(kr) \phi_\mathbf{k}\,,
\end{equation}
where $j_0(z) = \sin(z)/z$ is the zeroth-order spherical Bessel function.
Since both angular averaging and differentiation are linear operations, the radial profile $\phi(r)$ and its gradient $\phi'(r)$ inherit the Gaussian statistics of the primordial field $\phi(\mathbf{x})$. To facilitate the statistical analysis, we define the following auxiliary Gaussian variables:
\begin{equation}
	Y \equiv \phi(r)\,,\quad X \equiv r \phi'(r)\,.
\end{equation}
In our purely quadratic NG scenario, where the curvature perturbation is given by Eq.~(\ref{eq:purely_NG}), the linear compaction function reduces to
\begin{equation}
\label{eq:linear_compact_func}
    \mathcal{C}_\ell = -\frac{8A}{3} XY \,.
\end{equation}
The two–dimensional joint probability distribution of $X$ and $Y$ therefore follows a bivariate Gaussian distribution,
\begin{equation}
	\mathbb{P}(X, Y) = \frac{1}{2\pi \sqrt{\det(\mathbf{\Sigma})}} \exp\left[ -\frac{\mathbf{V}^T\mathbf{\Sigma}^{-1}\mathbf{V}}{2}\right]\,,\quad \mathbf{V}^T = (X,Y)\,, \quad \mathbf{\Sigma} = \begin{pmatrix}
		\Sigma_{XX} & \Sigma_{XY}\\
		\Sigma_{XY} & \Sigma_{YY}
	\end{pmatrix}\,.
\end{equation}
where the components of the covariance matrix are given by
\begin{align}
	\Sigma_{YY} (r) & \equiv \langle YY \rangle = \langle \phi(r)^2 \rangle = \int \mathrm{d}\ln k \, j_0^2(kr) \mathcal{P}_\phi(k)\,,\\
	\Sigma_{XY} (r) & \equiv \langle XY \rangle = \langle \phi(r) \cdot r\phi'(r) \rangle = \int \mathrm{d}\ln k \, (kr) j_0(kr) \frac{\mathrm{d}j_0}{\mathrm{d}z}(kr) \, \mathcal{P}_\phi(k)\,,\\
	\Sigma_{XX} (r) & \equiv \langle XX \rangle = \langle [r\phi'(r)]^2 \rangle = \int \mathrm{d}\ln k \, (kr)^2 \left[ \frac{\mathrm{d}j_0}{\mathrm{d}z}(kr) \right]^2 \mathcal{P}_\phi(k)\,.
\end{align}
The probability distribution of the linear compaction function $\mathcal{C}_\ell$ is obtained by marginalizing over the Gaussian variables $X$ and $Y$ using Eq.~(\ref{eq:linear_compact_func})
\begin{equation}
	\mathbb{P}(\mathcal{C}_\ell) = \int\mathrm{d}X \int\mathrm{d}Y\, \mathbb{P}(X, Y)\,\delta_D\left(\mathcal{C}_\ell + \frac{8A}{3} X Y\right) = \int\mathrm{d}Y \, \frac{3}{8|A Y|}\mathbb{P}\left(X\rightarrow-\frac{3\mathcal{C}_\ell}{8A Y}, Y\right),
\end{equation}
with $\delta_D$ denoting the Dirac delta function. Thanks to the purely quadratic nature of the NG, we derive the closed-form expression for the PDF of the linear compaction function
\begin{equation}
\label{eq:PDF_Bessel}
\mathbb{P}(\mathcal{C}_\ell) = \frac{3}{8\pi |A| \sqrt{\det(\mathbf{\Sigma})}} \exp\left( -\frac{3\Sigma_{XY} \mathcal{C}_\ell}{8A \det(\mathbf{\Sigma})} \right) K_0\left( \frac{3\sqrt{\Sigma_{XX}\Sigma_{YY}}}{8|A| \det(\mathbf{\Sigma})} \mathcal{C}_\ell \right),
\end{equation}
where $K_0$ is the modified Bessel function of the second kind of order zero.

Since PBH collapse requires the compaction function to exceed the threshold 
$\mathcal{C}_\ell \ge \mathcal{C}_{\ell,c}$, the large-$\mathcal{C}_\ell$ limit of 
Eq.~\eqref{eq:PDF_Bessel} provides an accurate description of the relevant probability tail, 
as confirmed numerically. 
Applying the asymptotic expansion of the modified Bessel function,
$K_0(z) \sim \sqrt{\pi/2z}\, e^{-z}$ for $z \gg 1$, the determinant factor in the normalization
cancels exactly, yielding a simplified analytic form for the tail:
\begin{equation}
\mathbb{P}(\mathcal{C}_\ell) \approx \frac{1}{\sqrt{2\pi \left( \frac{8}{3}|A| \sigma_X \sigma_Y \right) \mathcal{C}_\ell}} \exp\!\left[ - \frac{\mathcal{C}_\ell}{\frac{8}{3}|A| \sigma_X \sigma_Y \left( 1 - \mathrm{sgn}(A)\rho \right)} \right],
\label{eq:asymptotic_pdf}
\end{equation}
where the standard deviations and the dimensionless correlation coefficient are defined as
\begin{equation}
    \sigma_X \equiv \sqrt{\Sigma_{XX}}\,,\quad \sigma_Y \equiv \sqrt{\Sigma_{YY}}\,,\quad \rho \equiv \Sigma_{XY}/\sqrt{\Sigma_{XX}\Sigma_{YY}}\,.
\end{equation}

The asymptotic form in Eq.~(\ref{eq:asymptotic_pdf}) exhibits an exponential tail originating from the quadratic relation $\zeta \propto \phi^2$. Since the compaction function is proportional to the product of two Gaussian variables, $\mathcal{C}_\ell \propto XY$, the resulting probability distribution decays as $\sim \exp(-\mathcal{C}_\ell)$ rather than the much steeper Gaussian suppression $\sim \exp(-\mathcal{C}_\ell^2)$. In general, such a heavier tail would enhance the probability of large fluctuations and therefore exacerbate the PTA--PBH tension.

\begin{figure}[htbp!]
    \centering
    \includegraphics[width=.48\columnwidth]{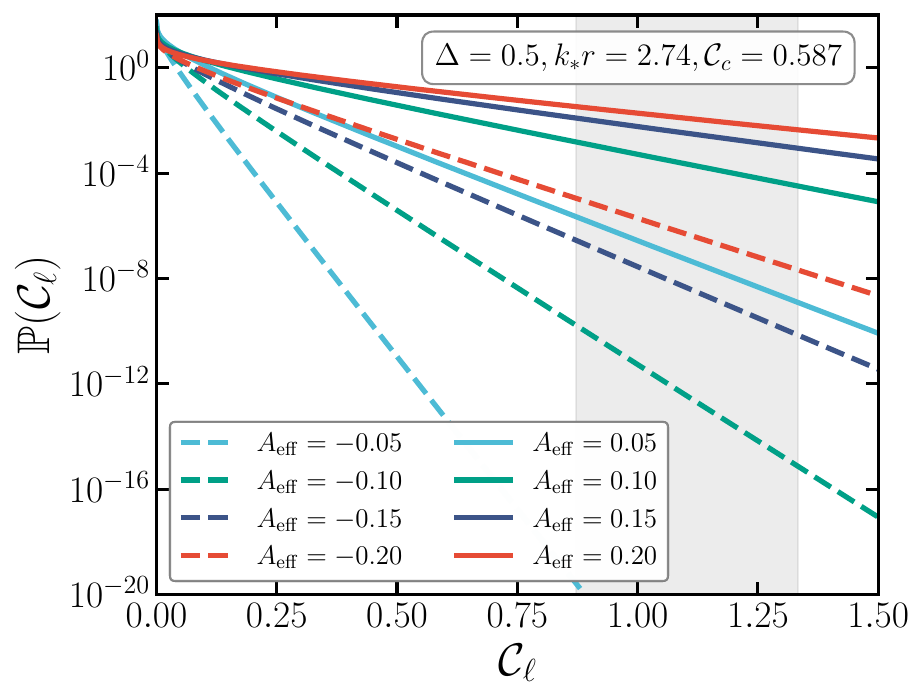}
    \hfill
    \includegraphics[width=.48\columnwidth]{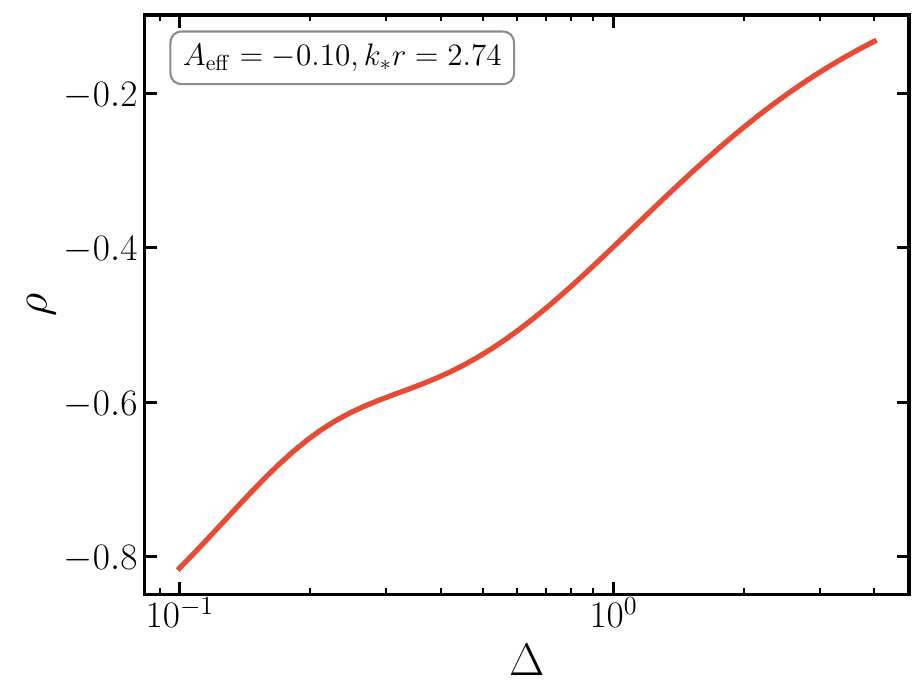}
    \caption{\label{fig:pcl_pdf_and_rho}%
    \textit{Left panel:} Probability density function of the linear compaction function $\mathcal{C}_\ell$ for a LN spectrum with fixed $\Delta = 0.5$, $k_\ast r = 2.74$, and $\mathcal{C}_c = 0.587$, shown for different values of the effective amplitude $A_{\rm eff}\equiv A\mathcal{A}_\phi$. Solid (dashed) lines correspond to $A_{\rm eff}>0$ ($A_{\rm eff}<0$). The shaded band marks the collapse region, bounded by the threshold $\mathcal{C}_{\ell,c}$ and the theoretical maximum $\mathcal{C}_{\ell,{\rm max}} = 4/3$ for Type I perturbations.
    \textit{Right panel:} Correlation coefficient $\rho$ as a function of the spectral width $\Delta$ for a LN spectrum with fixed $A_{\rm eff} = -0.1$ and $k_\ast r = 2.74$. Narrow spectra ($\Delta \to 0$) lead to strong anti-correlation $\rho \to -1$, suppressing the large-$\mathcal{C}_\ell$ tail via $\Lambda_{\rm tail} \propto (1+\rho)$. This results in an exponential suppression of the PBH formation probability.}
\end{figure}

However, in the present case the suppression scale of the tail is not determined solely by the variance $\sigma_{\mathcal{C}_{\ell}}^2$, as in the Gaussian case. Instead, Eq.~(\ref{eq:asymptotic_pdf}) shows that it is controlled by the combination $|A|\sigma_X\sigma_Y(1-\mathrm{sgn}(A)\rho)$, where $\rho$ quantifies the correlation between the smoothed field and its radial gradient. It is therefore convenient to define an effective tail scale,
\begin{equation}
    \Lambda_{\rm tail} \equiv \frac{8}{3}|A| \sigma_X \sigma_Y \left[1 - \mathrm{sgn}(A)\rho \right]\,,
\end{equation}
which directly governs the exponential decay of the compaction distribution. As shown in the left panel of Fig.~\ref{fig:pcl_pdf_and_rho}, different values of the non-Gaussian coefficient $A$, together with the variances $\sigma_X$ and $\sigma_Y$, modify $\Lambda_{\rm tail}$, leading to visibly different tail slopes: negative $A$ (dashed curves) results in much stronger suppression, while positive $A$ (solid curves) produces comparatively enhanced tails.
The key behavior arises for $A<0$, for which one has $\Lambda_{\rm tail} \propto (1+\rho)$, so that in the strong anticorrelation limit $\rho\to -1$ the probability of large compaction fluctuations is exponentially suppressed.
Accordingly, the correlation coefficient $\rho$ exhibits a strong dependence on the width of the source power spectrum, as illustrated in the right panel of Fig.~\ref{fig:pcl_pdf_and_rho} for a representative LN spectrum.
\footnote{For a spectrum sharply localized around a characteristic wave number, the covariance integrals are dominated by a narrow range of modes, so that the field amplitude $Y = \phi(r)$ and its radial gradient $X = r\phi'(r)$ are nearly deterministically related, yielding $|\rho| \to 1$. For a broad spectrum with extended support in wave number space, $\Sigma_{XY}$ receives contributions from modes with different phases. Owing to the oscillatory kernel $k r j_0(kr) j_0'(kr)$, these contributions partially cancel, whereas $\Sigma_{XX}$ and $\Sigma_{YY}$ remain positive-definite. This results in an efficient decorrelation between $X$ and $Y$, suppressing $|\Sigma_{XY}|$ relative to $\sqrt{\Sigma_{XX}\Sigma_{YY}}$, and hence $|\rho| \ll 1$.}
In particular, as the spectrum becomes sharply peaked ($\Delta \to 0$), or equivalently in the monochromatic limit $\mathcal{P}_\phi^\delta(k) = \mathcal{A}_\phi\, \delta_D(\ln k - \ln k_\ast)$, the correlation coefficient approaches $\rho \to -1$. This drives $\Lambda_{\rm tail} \to 0$, leading to $\mathbb{P}(\mathcal{C}_\ell) \to 0$ in the large-$\mathcal{C}_\ell$ regime.
By contrast, for $A>0$, one instead has $\Lambda_{\rm tail} \propto (1-\rho)$, which increases the effective tail scale and enhances the probability of large fluctuations.

This suppression for $A<0$ can be understood intuitively. When the smoothed field and its radial gradient are strongly anti-correlated, typical field profiles correspond to a central peak that rapidly decreases with radius. Such configurations tend to produce negative contributions to the compaction function, making it difficult to generate the large positive values required for gravitational collapse. Consequently, the probability of exceeding the collapse threshold $\mathcal{C}_{\ell,c}$ becomes strongly suppressed.

\subsection{PBH mass spectrum and abundance}

With the analytical probability distribution $\mathbb{P}(\mathcal{C}_\ell)$ established, we proceed to evaluate the abundance of PBHs in the presence of quadratic NG.
We consider PBH formation during the radiation-dominated era, where sufficiently large density perturbations on a comoving smoothing scale $r$ re-enter the horizon. The corresponding horizon mass at this crossing time is given by~\cite{Ando:2018qdb}
\begin{equation}
\label{eq:horizon_mass}
    M_H(r) = \rho_r\frac{4\pi}{3}H_r^{-3} \simeq \frac{M_\mathrm{eq}}{\sqrt{2}}\left(\frac{g_{\ast,\text{eq}}}{g_{\ast,r}}\right)^{1/6}\left(k_\mathrm{eq}r\right)^2\,.
\end{equation}
Here, $M_{\text{eq}} \approx 7 \times 10^{50}\text{ g}$ denotes the horizon mass at the matter-radiation equality time, and $k_\text{eq} = a_\text{eq}H_\text{eq}$ is the associated comoving wave number. The parameters $g_{\ast,r} \approx 106.75$ and $g_{\ast,\text{eq}} \approx 3.36$ represent the effective number of relativistic degrees of freedom for the radiation energy density at the time of PBH formation and at the equality time, respectively.
Following the scaling law of critical collapse~\cite{Choptuik:1992jv,Evans:1994pj,Niemeyer:1997mt}, the mass of the resulting PBH is related to the horizon mass $M_H$ by
\begin{equation}
\label{eq:scaling_law}
    M_\mathrm{PBH}(\mathcal{C}_\ell) = \mathcal{K} M_H \left(\mathcal{C}_\ell-\frac{3}{8}\mathcal{C}_\ell^2-\mathcal{C}_c\right)^\gamma\,,
\end{equation}
where the coefficient $\mathcal{K}$ depends mildly on the peak profile and is not yet precisely determined; following previous works we adopt $\mathcal{K}\simeq 4$. The critical exponent $\gamma$ depends on the equation of state and takes $\gamma\simeq0.36$ in a radiation-dominated universe.
For the collapse threshold and smoothing scale entering the compaction formalism, we adopt a pragmatic approach. The relation between the smoothing scale $r$ and the characteristic scale $k_\ast$ depends on the shape of the perturbation profile and is not universal. 
For Gaussian perturbations, numerical simulations indicate that the overdensity typically peaks at $r \simeq 2.74/k_\ast$ for a monochromatic spectrum, and at $r \simeq 4.49/k_\ast$ for a nearly scale-invariant spectrum~\cite{Musco:2008hv,Musco:2020jjb,Germani:2018jgr}. However, these results rely on Gaussian statistics and do not directly extend to the strongly non-Gaussian regime considered here. 
We therefore treat both $k_\ast r$ and the collapse threshold $\mathcal{C}_c$ as phenomenological parameters and explore their impact on the PBH abundance below.

In the extended Press-Schechter formalism, the fraction of PBHs relative to the total energy density at the time of formation is given by
\begin{equation}
	\beta(r) \equiv \left.\frac{\rho_{\mathrm{PBH}}}{\rho_r}\right|_\text{formation} =  \int_{\mathcal{C}_{\ell,c}}^{4/3}\mathrm{d}\mathcal{C}_\ell\, \frac{M_\mathrm{PBH}(\mathcal{C}_\ell)}{M_H}\mathbb{P}(\mathcal{C}_\ell)\,,
\end{equation}
where the lower bound $\mathcal{C}_{\ell,c}$ is the linear compaction threshold corresponding to the nonlinear threshold $\mathcal{C}_c$, obtained by inverting Eq.~(\ref{eq:non-linear-Com}).
Crucially, unlike the conventional density contrast integrated to infinity, PBH formation from Type I perturbations is bounded within a finite collapse range, $\mathcal{C}_{\ell,c} \le \mathcal{C}_\ell \le 4/3$.

The present-day mass distribution function is defined as~\cite{Kitajima:2021fpq}
\begin{equation}
    f_\mathrm{PBH}(M) \equiv \frac{\rho_{\mathrm{PBH},0}}{\rho_{\mathrm{DM},0}} \simeq \frac{\Omega_{\mathrm{m},0}}{\Omega_{\mathrm{DM},0}}\frac{1}{k_\mathrm{eq} r}\beta(M)\,,
\end{equation}
where $\Omega_{\mathrm{m},0} \approx 0.31$ and $\Omega_{\mathrm{DM},0} \approx 0.26$ are the current matter and DM density parameters, respectively~\cite{Planck:2018vyg}, and
\begin{equation}
    \beta(M) = \left|\frac{\mathrm{d}\ln M}{\mathrm{d}\mathcal{C}_\ell}\right|^{-1}\frac{M(\mathcal{C}_\ell)}{M_H}\mathbb{P}(\mathcal{C}_\ell) = \frac{\mathcal{K}(\mathcal{C}_\ell-\frac{3}{8}\mathcal{C}_\ell^2-\mathcal{C}_c)^{\gamma+1}}{\gamma \left(1-\frac{3}{4}\mathcal{C}_\ell\right)} \mathbb{P}(\mathcal{C}_\ell)\,.
\end{equation}
Here, the linear compaction function is determined by the PBH mass $M$ as
\begin{equation}
    \mathcal{C}_\ell = \frac{4}{3}\left(1-\sqrt{1-\frac{3}{2}\mathcal{C}_M}\right)\,,\quad \mathcal{C}_M = \mathcal{C}_c + \left(\frac{M}{\mathcal{K} M_H}\right)^{1/\gamma}.
\end{equation}
The total PBH abundance is then obtained by integrating over the mass spectrum,
\begin{equation}
    f_{\mathrm{PBH}}^{\mathrm{tot}} \equiv \frac{\rho_{\mathrm{PBH},0}}{\rho_{\mathrm{DM},0}}
    = \int_0^{M_\mathrm{max}}\mathrm{d}\ln M\, f_\mathrm{PBH}(M)\,,
\end{equation}
where $M_\mathrm{max}$ is the maximum PBH mass, obtained by evaluating the scaling law in Eq.~(\ref{eq:scaling_law}) at $\mathcal{C}_\ell = 4/3$.
It is worth noting that since the PBH mass $M(\mathcal{C}_\ell)$ varies quadratically with the compaction function near its maximum, the Jacobian $|\mathrm{d}\ln M/\mathrm{d}\mathcal{C}_\ell|^{-1}$ exhibits a divergence of the form $(M_\mathrm{max}-M)^{-1/2}$ as $M \to M_\mathrm{max}$. Nevertheless, this divergence is integrable, guaranteeing that the total PBH abundance  $f_{\mathrm{PBH}}^{\mathrm{tot}}$ remains well-defined and finite~\cite{Abe:2022xur}.

\begin{figure}[htbp!]
    \centering
    \includegraphics[width=.48\columnwidth]{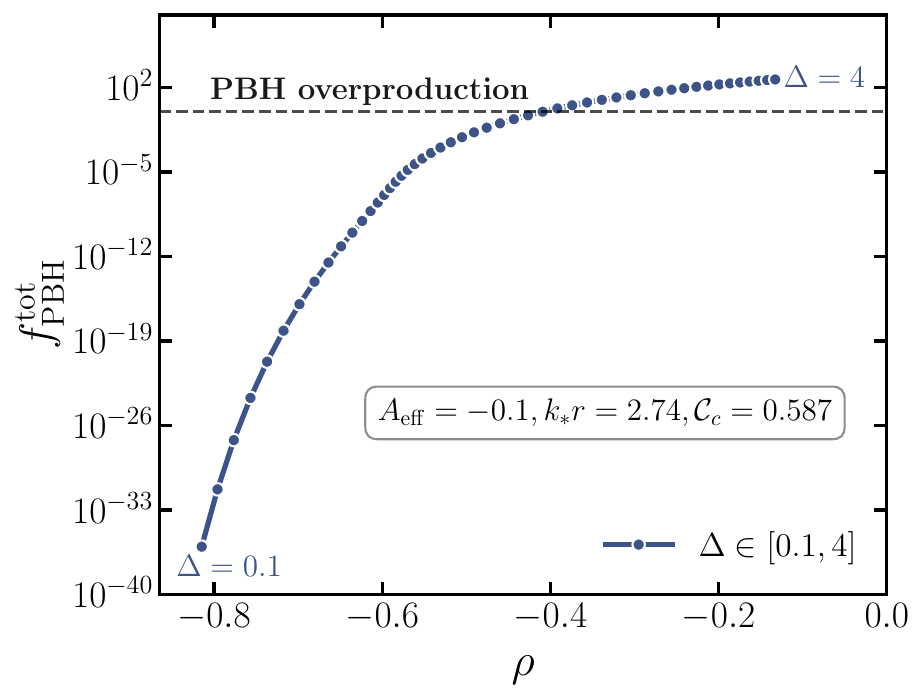}%
    \hfill
    \includegraphics[width=.48\columnwidth]{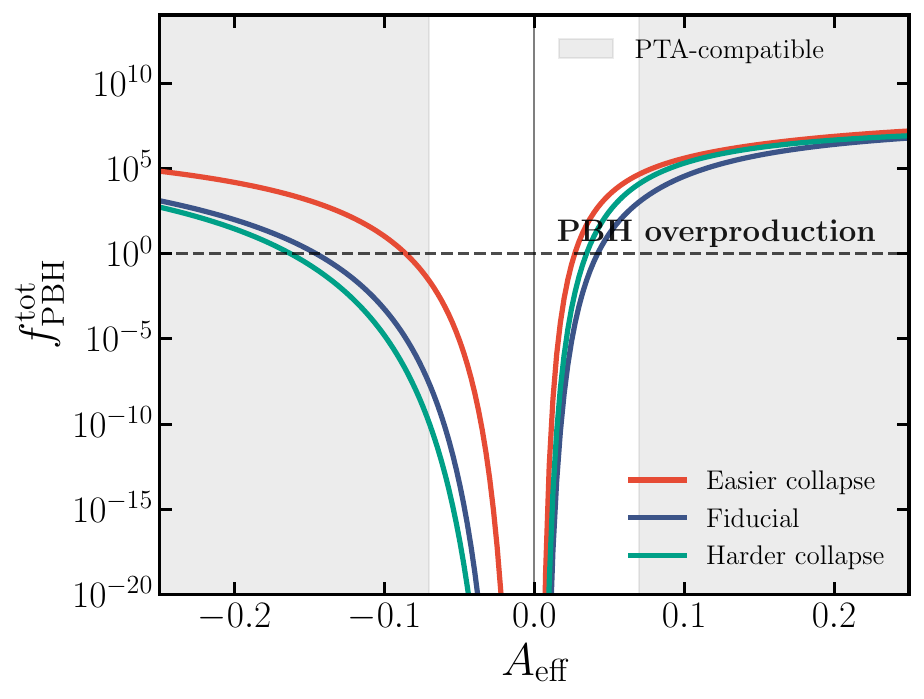}
    \caption{\label{fig:f_PBH_tot}%
    \textit{Left panel:} Total PBH abundance $f_{\rm PBH}^{\rm tot}$ as a function of the correlation coefficient $\rho=\rho(\Delta)$, obtained by varying the width parameter $\Delta\in[0.1,4]$ in the LN model in Eq.~(\ref{eq:LN_spectrum}), while fixing the effective amplitude $A_{\rm eff} = A\mathcal{A}_\phi$, the smoothing scale $r$, the collapse threshold $\mathcal{C}_c$, and the characteristic scale $k_\ast = 3\times10^{6}\,\mathrm{Mpc}^{-1}$. 
    As $\Delta$ decreases, the spectrum becomes increasingly peaked, driving $\rho$ toward $-1$ and thereby exponentially suppressing PBH formation, in accordance with the scaling $\Lambda_{\rm tail}\propto (1+\rho)$.
    \textit{Right panel:} Total PBH abundance $f_{\rm PBH}^{\rm tot}$ as a function of the effective amplitude $A_{\rm eff}$, computed using the same LN spectrum with fixed $\Delta=0.5$ and $k_\ast = 3\times10^{6}\,\mathrm{Mpc}^{-1}$. 
    The grey shaded band indicates the PTA-preferred region inferred from the stochastic GW signal, while the horizontal dashed line marks $f_{\rm PBH}^{\rm tot}=1$, above which PBHs are overproduced. Three representative collapse prescriptions are shown: an easier collapse case $(k_\ast r=3.0,\, \mathcal{C}_c=0.4)$, a fiducial setup $(k_\ast r=2.74,\, \mathcal{C}_c=0.587)$, and a harder collapse case $(k_\ast r=1.5,\, \mathcal{C}_c=0.65)$.}
\end{figure}
Since PBH formation is dominated by rare large fluctuations, the abundance is controlled by the exponential tail of the compaction distribution. Using the asymptotic form derived in Eq.~(\ref{eq:asymptotic_pdf}), one finds
\begin{equation}
f_{\rm PBH} \sim \exp\!\left(-\frac{\mathcal{C}_{\ell,c}}{\Lambda_{\rm tail}}\right).
\end{equation}
This expression makes explicit that the PBH abundance is exponentially sensitive not only to the amplitude of fluctuations, but also to the correlation coefficient and the collapse threshold. In particular, for negative quadratic non-Gaussian parameter ($A<0$), one has $\Lambda_{\rm tail} \propto (1+\rho)$, so that the limit $\rho \to -1$ leads to a vanishing tail scale and hence an exponential suppression of PBH formation.

To isolate this effect, in the left panel of Fig.~\ref{fig:f_PBH_tot} we vary the width parameter $\Delta$ of the LN spectrum while fixing the effective amplitude $A_{\rm eff} = A \mathcal{A}_\phi$, the smoothing scale $r$ and the collapse threshold $\mathcal{C}_c$.
Since $\Delta$ characterizes the spectral width, this procedure effectively induces a scan over the correlation coefficient $\rho$, whose dependence on $\Delta$ is shown in the right panel of Fig.~\ref{fig:pcl_pdf_and_rho}. We find that narrower spectra (smaller $\Delta$) drive $\rho$ toward $-1$, resulting in a dramatic suppression of the PBH abundance over many orders of magnitude. This highlights that the correlation coefficient plays a central role in governing the tail of the compaction distribution.

To demonstrate how the PTA signal can be reconciled with PBH constraints, we show in the right panel of Fig.~\ref{fig:f_PBH_tot} the total PBH abundance $f_{\rm PBH}^{\rm tot}$ as a function of the amplitude $A_{\rm eff}$. 
The grey shaded region indicates the PTA-preferred range.
To account for the theoretical uncertainty in PBH formation, we consider three representative collapse prescriptions, characterized by different combinations of the smoothing scale $r$ and the threshold $\mathcal{C}_c$: an easier-collapse case ($k_\ast r=3.0,\,\mathcal{C}_c=0.4$), a fiducial setup ($k_\ast r=2.74,\,\mathcal{C}_c=0.587$), and a harder-collapse case ($k_\ast r=1.5,\,\mathcal{C}_c=0.65$). 
These choices effectively span a range of collapse efficiencies.

In the positive branch ($A_{\rm eff}>0$), the PBH abundance is strongly enhanced due to the broadening of the exponential tail in the PDF.
As a result, within the PTA-preferred amplitude range, all three prescriptions predict $f_{\rm PBH}^{\rm tot} \gg 1$, indicating severe overproduction of PBHs. 
This behavior is relatively insensitive to the choice of collapse prescription, implying that the tension between PTA and PBH constraints persists in the positive non-Gaussian case.

In contrast, the negative branch ($A_{\rm eff}<0$) exhibits qualitatively different behavior.
In this regime, the abundance is exponentially suppressed, following $f_{\rm PBH} \sim \exp(-\mathcal{C}_{\ell,c}/\Lambda_{\rm tail})$, where the effective tail scale scales as $\Lambda_{\rm tail} \propto (1+\rho)$.
As the correlation coefficient approaches $\rho \to -1$, the tail of the compaction distribution is strongly suppressed, leading to an exponential reduction of the PBH abundance.
As a result, the predicted abundance becomes highly sensitive to the collapse prescription.
While the easier-collapse case can still lead to overproduction within the PTA window, the fiducial and harder-collapse setups yield $f_{\rm PBH}^{\rm tot}\lesssim 1$, thereby allowing the PTA signal to be accommodated without violating PBH bounds.

These results demonstrate that purely quadratic NG can substantially reduce the PBH abundance through correlation-induced suppression of the tail of the compaction function distribution. However, the quantitative efficiency of this mechanism depends sensitively on the collapse prescription, particularly on the smoothing scale and the collapse threshold. In the following section, we confront this framework with the NANOGrav 15-year data and identify the parameter regions consistent with both the observed SGWB and current PBH constraints.

\section{SIGWs and Observational Implications}
\label{sec:SIGW}

Alongside PBH formation, the enhanced scalar perturbations re-entering the horizon during the radiation-dominated era inevitably generate a stochastic background of SIGWs. In this section, we explore the observational consequences of this unified framework across different scales. We first compute the SIGW spectrum and confront it with PTA observations, assessing the extent to which the PTA–PBH tension can be alleviated. We then show that the same mechanism, when extended to smaller scales, naturally gives rise to asteroid-mass PBH DM accompanied by a high-frequency GW signal within the sensitivity of future space-based interferometers.

\subsection{SIGW spectrum from quadratic perturbations}
The large curvature perturbations on small scales responsible for PBH formation also act as a source for SIGWs. Through second-order mode coupling, these fluctuations induce tensor modes upon horizon reentry during the radiation-dominated epoch.
The generation of SIGWs acts most efficiently around horizon crossing and rapidly becomes negligible shortly after reentry due to the oscillation and decay of the sourcing scalar perturbations.

At a later time when the GW density fraction becomes constant, the energy density parameter can be expressed analytically in terms of the primordial curvature power spectrum $\mathcal{P}_\zeta$ as~\cite{Kohri:2018awv,Espinosa:2018eve,Inomata:2019yww}
\begin{equation}
\Omega^{\rm RD}_{\rm GW}(k) = \frac{1}{2} \int_{-1}^{1} \mathrm{d}x \int_{1}^{\infty} \mathrm{d}y \, 
\mathcal{T}(x,y)\mathcal{P}_{\zeta}\left( \frac{y - x}{2}k \right) 
\mathcal{P}_{\zeta}\left( \frac{x + y}{2}k \right) , 
\end{equation}
where the transfer function
\begin{equation}
\mathcal{T}(x,y) = \frac{(x^2 + y^2 - 6)^2 (x^2 - 1)^2 (y^2 - 1)^2}{(x - y)^8 (x + y)^8} 
\left\{
\left[ x^2 - y^2 + \frac{x^2 + y^2 - 6}{2} \ln\left| \frac{y^2 - 3}{x^2 - 3} \right| \right]^2 
+ \frac{\pi^2 (x^2 + y^2 - 6)^2}{4} \Theta(y - \sqrt{3})
\right\},
\end{equation}
and $\Theta$ represents the Heaviside step function.
Deep inside the horizon, the source term decays rapidly, allowing the induced GWs to propagate freely as radiation with energy density $\rho_{\mathrm{GW}} \propto a^{-4}$. The present-day spectrum of SIGWs is therefore given by
\begin{equation}
    h^2 \Omega_{\rm GW}(k) = h^2 \Omega_{r,0} 
\left( \frac{g_{\ast,r}}{g_{\ast,0}} \right)^{-1/3} \Omega^{\rm RD}_{\rm GW}(k) \,,
\end{equation}
where $h^2\Omega_{r,0}\simeq 4.2 \times 10^{-5}$ is the current radiation abundance~\cite{Planck:2018vyg}, and $g_{\ast,0}$ and $g_{\ast,r}$ denote the effective relativistic degrees of freedom today and at the time of SIGW production, respectively.

The corresponding GW frequency is related to the comoving wave number by
\begin{equation}
f \simeq 1.6\, \mathrm{nHz} \left( \frac{k}{10^6\, \mathrm{Mpc}^{-1}} \right).
\label{eq:freq_k_relation}
\end{equation}
This relation establishes a direct correspondence between the GW frequency and the characteristic scale of the primordial perturbations, and hence, via Eq.~(\ref{eq:horizon_mass}), to the PBH mass. As a result, PTA frequencies probe perturbations associated with stellar-mass PBHs, while space-based interferometers such as LISA are sensitive to much smaller scales corresponding to PBHs in the asteroid-mass window.

\subsection{Constraints from PTA data}

We now confront the purely quadratic non-Gaussian scenario with the NANOGrav 15-year dataset to assess whether it can alleviate the PTA–PBH tension. We employ Bayesian inference utilizing the Python package \texttt{PTArcade}~\cite{Mitridate:2023oar} in the \emph{ceffyl} mode~\cite{Lamb:2023jls} to constrain the model parameters, aiming to identify a viable parameter space that accommodates the GW signal while maintaining the PBH abundance within safe observational limits.

We model the curvature power spectrum using a BPL parametrization, which naturally captures the robust $k^3$ infrared scaling characteristic of tachyonic amplification and provides an accurate description of the low-frequency GW signal in the PTA band. This form is also well motivated in thermal inflation~\cite{Dimopoulos:2019wew,Lewicki:2021xku} and hybrid inflation scenarios~\cite{Fonseca:2010nk,Gong:2010zf,Lyth:2011kj,Lyth:2012yp}.
Since current PTA data are largely insensitive to the high-frequency tail, the UV spectral index $\beta$ and the smoothness parameter $s$ exhibit a strong degeneracy in determining the effective spectral width. To avoid this degeneracy, we fix $s = 1$ and treat $\beta$ as an effective proxy for the spectral width.

The base parameters and their corresponding priors used in our numerical analysis are summarized in Table~\ref{tab:Prior_and_posterior}. We use the effective amplitude parameter $A_{\rm eff} = A\mathcal{A}_\phi$ to improve the convergence of the Markov Chain Monte Carlo (MCMC) chains, as the non-Gaussian parameter $A$ and the primordial amplitude $\mathcal{A}_\phi$ are strongly degenerate in the likelihood space. In practice, the PTA data are primarily sensitive to their product rather than to each parameter individually.
For the priors, we adopt $\log_{10} |A_{\rm eff}| \in [-5, 0.5]$, which covers the relevant range of PBH abundances. The characteristic frequency is taken to lie within $\log_{10}(f_\ast/\mathrm{Hz}) \in [-10, -6]$, fully encompassing the sensitivity band of the NANOGrav 15-year dataset. Here, $f_\ast$ denotes the frequency corresponding to the characteristic scale $k_\ast$ via Eq.~(\ref{eq:freq_k_relation}) and should be distinguished from the peak frequency of the induced GW spectrum.
To assess the impact of the spectral shape on the GW signal, we further allow the UV slope to vary within $\beta \in [1, 30]$.

\begin{table}[h]
\caption{\label{tab:Prior_and_posterior}%
Prior distributions of the model parameters for the BPL spectrum and the posterior mean values with $1\sigma$ uncertainties obtained from the fit to the NANOGrav 15-year dataset.}
\begin{ruledtabular}
\begin{tabular}{cccc}
Parameters & Description & Prior & Posterior mean\\
\colrule
$\log_{10}|A_{\rm eff}|$ & effective amplitude & log-uniform [-5, 0.5] & $-0.59^{+0.25}_{-0.27}$\\
$\log_{10}\left(f_\ast/{\rm Hz}\right)$ & characteristic frequency & log-uniform [-10, -6] & $-7.33^{+0.41}_{-0.43}$\\
$\log_{10}\beta$ & UV spectral index & log-uniform [0, 1.48] & $0.73^{+0.51}_{-0.50}$\\
\end{tabular}
\end{ruledtabular}
\end{table}

\begin{figure}[htbp!]
    \centering
    \includegraphics[width=.56\columnwidth]{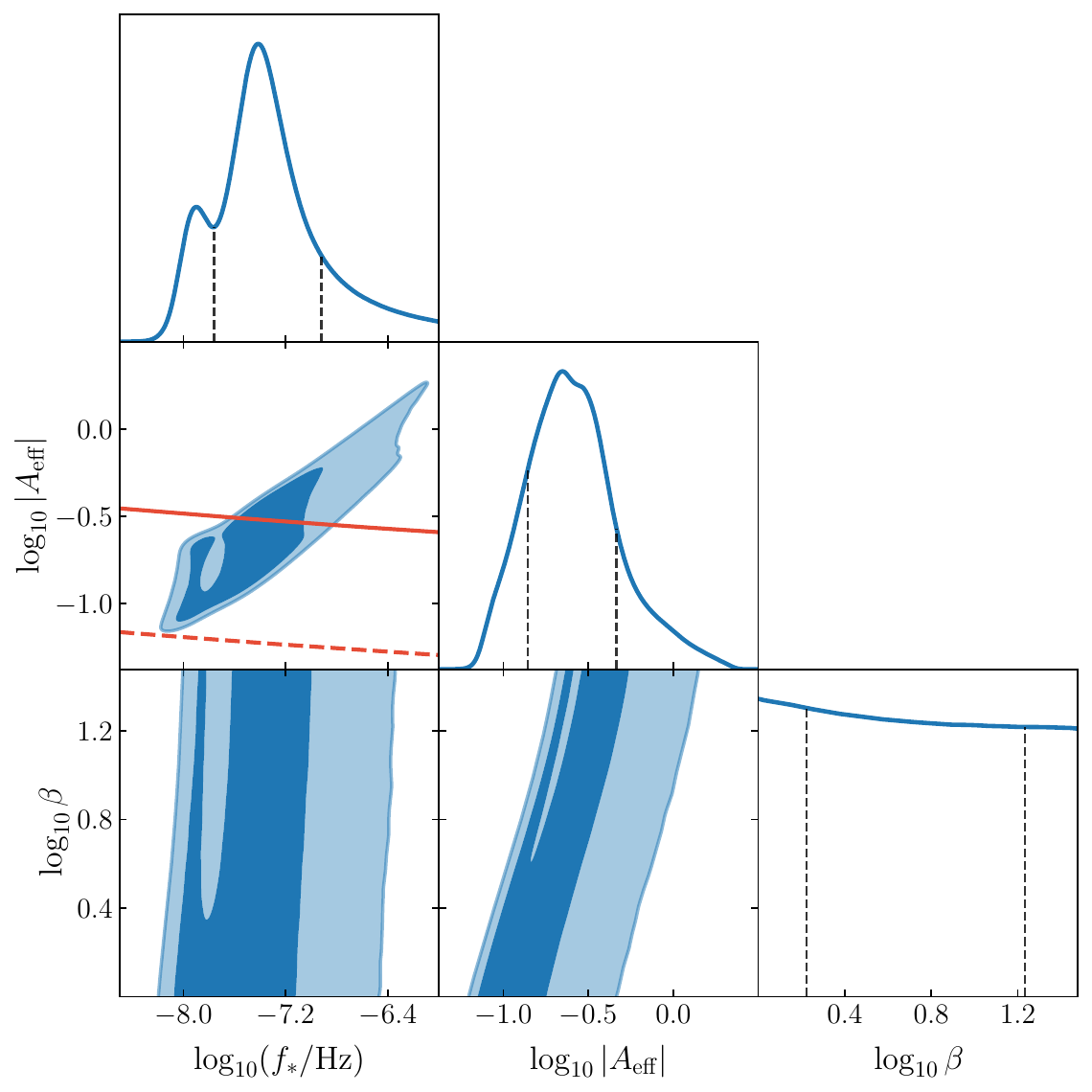}
    \caption{\label{fig:posterior}%
    Posterior distributions for the parameters of the BPL model inferred from the NANOGrav 15-year dataset. The dark and light blue contours denote the 68\% and 95\% credible levels, respectively. In the diagonal panels, the dashed vertical lines indicate the $68\%$ credible level. The solid and dashed red curves delineate the $f_{\mathrm{PBH}}^{\mathrm{tot}} = 1$ contour for the negative ($A_{\mathrm{eff}}<0$) and positive ($A_{\mathrm{eff}}>0$) non-Gaussian branches, respectively, with the region above the upper curve excluded by observations. The band enclosed by these two red lines defines the viable parameter window for $|A_{\mathrm{eff}}|$, which successfully accommodates the PTA signal while simultaneously generating a significant abundance of PBH DM.
    }
\end{figure}

The resulting posterior distributions of the model parameters, obtained under the assumption of a BPL spectrum, are shown in Fig.~\ref{fig:posterior}. For the PBH abundance calculation, we adopt a fiducial setup ($k_\ast r=2.74,\,\mathcal{C}_c=0.587$), and the qualitative results are robust against this choice.
As inferred from the posterior distributions in Fig.~\ref{fig:posterior}, the positive non-Gaussian branch ($A_{\rm eff}>0$, dashed red curve) is incompatible with PBH constraints within the 95\% credible level preferred by the PTA data, as it generically leads to excessive PBH production due to the enhanced tail of the compaction distribution.
By contrast, the negative non-Gaussian branch ($A_{\rm eff}<0$, solid red curve) admits viable solutions. In this case, scenarios avoiding PBH overproduction lie within the 68\%–-95\% credible level of the PTA signal. Within our setup, this compatibility is realized within a restricted range of the effective amplitude, $-0.31 \lesssim A_{\mathrm{eff}} \lesssim -0.07$.
In this regime, the model can reproduce the PTA signal while generating a significant, yet observationally allowed, PBH abundance. For more negative values of $A_{\mathrm{eff}}$ ($A_{\mathrm{eff}} \lesssim -0.31$), PBH overproduction renders the scenario experimentally excluded. Conversely, for less negative values of $A_{\mathrm{eff}}$ ($A_{\mathrm{eff}} \gtrsim -0.07$), the induced GW signal becomes too weak to account for the PTA excess, reflecting the strong dependence of the GW energy density on the effective amplitude.

\begin{figure}[htbp!]
    \centering
    \includegraphics[width=.56\columnwidth]{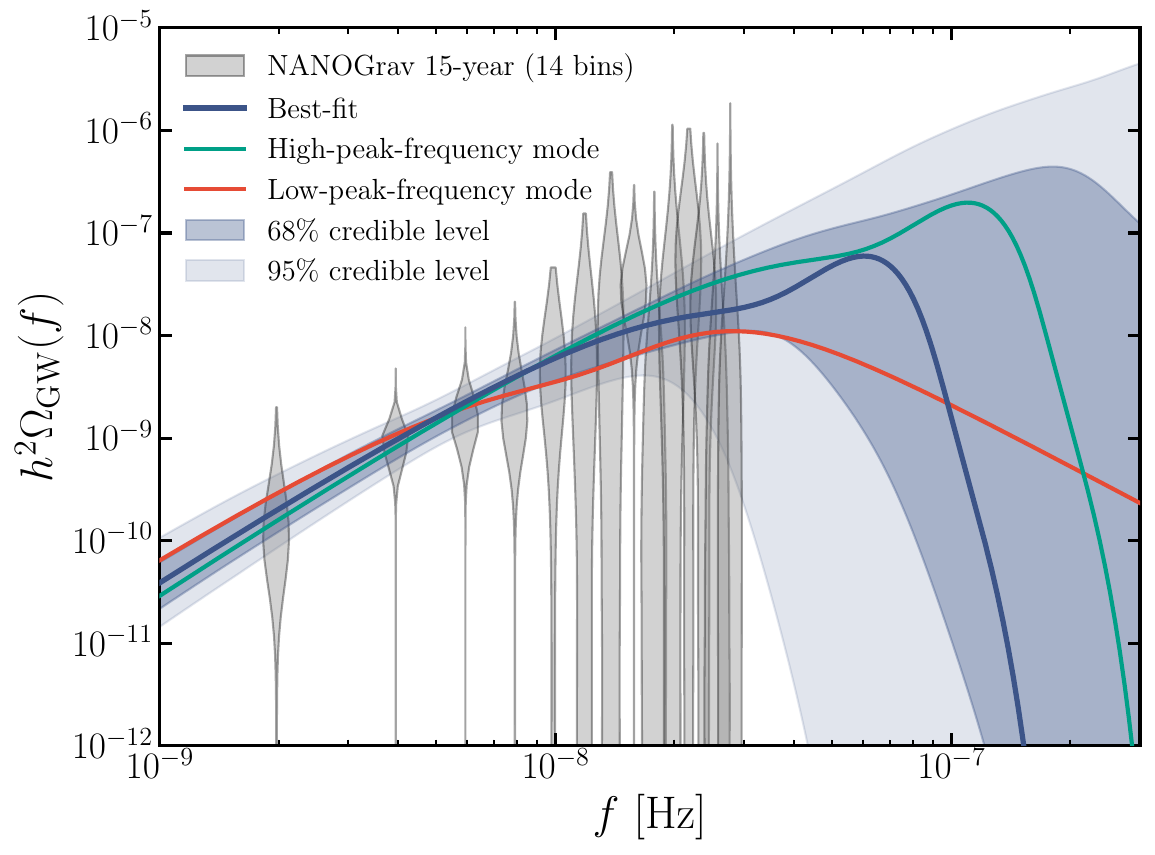}
    \caption{\label{fig:bimodality_plot}%
    Comparison of the predicted SIGW energy density spectra with the first 14 frequency bins of the NANOGrav 15-year free-spectrum dataset, shown as grey violin plots. The blue curve represents the best-fit spectrum, while the darker and lighter blue shaded bands denote the $68\%$ and $95\%$ posterior credible levels, respectively. The colored curves illustrate two representative regimes associated with the bimodal posterior of the characteristic frequency: the green curve (High-peak-frequency mode, $\log_{10}(f_\ast/\mathrm{Hz}) \sim -7.4$) mainly fits the lowest-frequency bins via the $f^3$ infrared scaling, while the red curve (Low-peak-frequency mode, $\log_{10}(f_\ast/\mathrm{Hz}) \sim -8.0$) better matches the intermediate-frequency bins around $10^{-8}\,\mathrm{Hz}$ due to the transient flattening of the BPL spectrum.
    }
\end{figure}

A notable feature of our Bayesian analysis is the bimodality in the one-dimensional marginalized posterior for the characteristic frequency, with a primary mode at $\log_{10}(f_\ast/\mathrm{Hz}) \sim -7.4$ and a secondary mode at $\sim -8.0$. As illustrated in Fig.~\ref{fig:bimodality_plot}, this reflects two distinct regions in parameter space that provide comparably good fits to the data. In the high-peak-frequency case (green curve), the SGWB peak is pushed to $\sim 10^{-7}$ Hz, well outside the optimal NANOGrav band. Consequently, the spectrum mainly accommodates the lowest-frequency PTA bins but deviates at higher frequencies. Conversely, the low-peak-frequency case (red curve) utilizes a characteristic feature of the BPL spectrum: a transient flattening of the $f^3$ infrared growth just before the peak. While this specific spectral shape slightly compromises the fit at lower frequencies, it enables the curve to tightly match the central data bins around $f \sim 10^{-8}$ Hz.

Furthermore, the posterior for the UV spectral index $\beta$ remains entirely flat, indicating it is unconstrained by the current PTA dataset. This lack of sensitivity arises because $\beta$ exclusively governs the descending part of the SIGW spectrum. Since the most sensitive NANOGrav 15-year frequency bins capture only the infrared rising phase and the peak, the high-frequency regime falls well outside the observational window. Consequently, varying $\beta$ has a negligible effect on the theoretical spectrum in the relevant frequency range, leaving the likelihood insensitive to this parameter.

\subsection{PBH DM and prospects for future detectors}

Apart from the PTA-relevant regime, the abundance of stellar-mass PBHs is already strongly constrained by the LVK~\cite{LIGOScientific:2018mvr,LIGOScientific:2020ibl,KAGRA:2021vkt,LIGOScientific:2025slb} GW detections of binary mergers and the CMB limits~\cite{Ali-Haimoud:2016mbv,Poulin:2017bwe,Serpico:2020ehh} associated with PBH accretion. We therefore turn to a distinct mass range in which PBHs can still constitute all of the DM.

\begin{figure}[htbp!]
    \centering
    \includegraphics[width=.48\columnwidth]{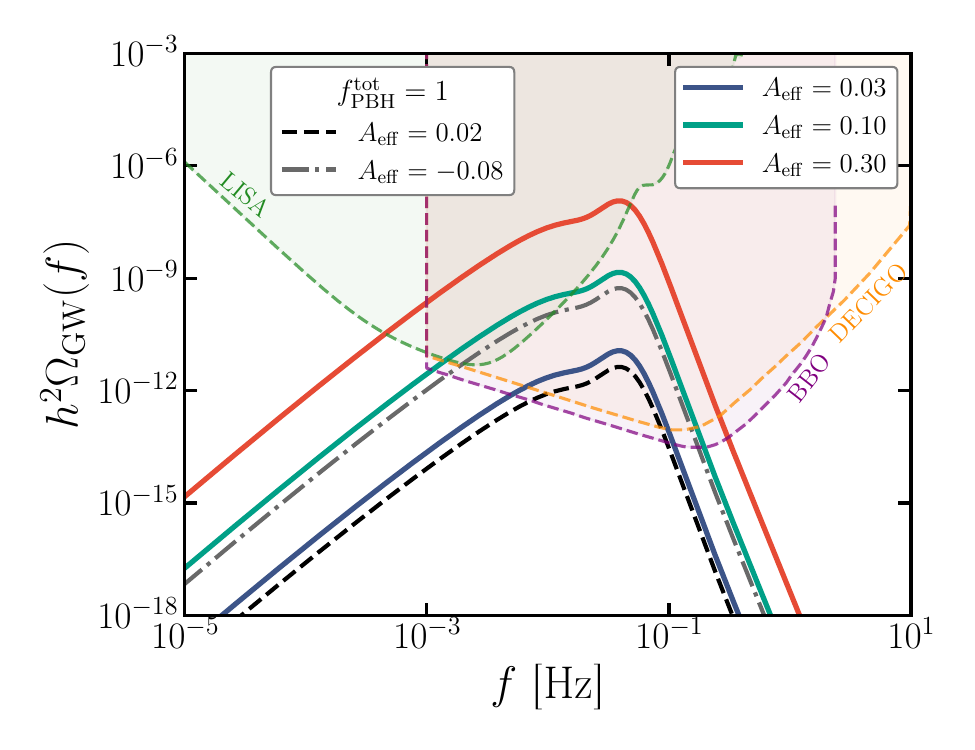}
    \hfill
    \includegraphics[width=.48\columnwidth]{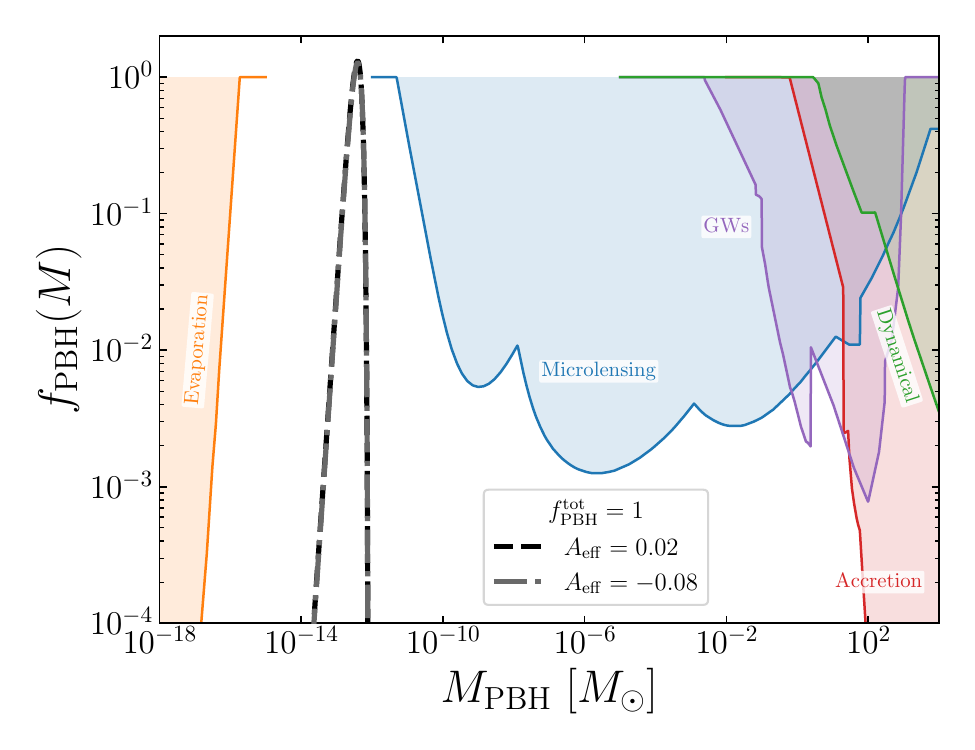}
    \caption{\label{fig:GW_and_fPBH}%
    Scalar-induced gravitational-wave (SIGW) spectra and corresponding PBH mass distribution.
    \textit{Left panel:} SIGW spectra for a BPL primordial power spectrum with varying effective amplitudes $A_{\mathrm{eff}}$. The solid curves show representative values $A_{\mathrm{eff}} = 0.03$ (blue), $0.10$ (green), and $0.30$ (red). Dashed ($A_{\mathrm{eff}}=0.02$) and dash-dotted ($A_{\mathrm{eff}}=-0.08$) curves correspond to $f_{\mathrm{PBH}}^{\mathrm{tot}} = 1$. Shaded regions indicate the projected sensitivities of future detectors.
    \textit{Right panel:} Extended PBH mass distributions in the asteroid-mass window are shown, where $f_{\rm PBH}(M) = \mathrm{d} f_{\rm PBH}^{\mathrm{tot}}/\mathrm{d}\ln M$ denotes the PBH abundance per logarithmic mass interval and can locally exceed unity.
    The dashed and dash-dotted curves correspond to the same values of $A_{\rm eff}$ as in the left panel, with $f_{\rm PBH}^{\mathrm{tot}} = 1$. Colored regions denote existing observational bounds for monochromatic PBHs, obtained using the \texttt{Python} code \texttt{PBHbounds}~\cite{10.5281/zenodo.3538999}, including Hawking evaporation (orange), microlensing experiments (blue), merger rates from GWs (purple), dynamical effects (green), and CMB accretion (red), and are shown for reference only.}
\end{figure}

Lighter PBHs lose mass and evaporate via Hawking radiation, emitting high-energy particles (such as gamma rays and electron–positron pairs) into the cosmic medium. This leads to stringent constraints from BBN~\cite{Carr:2009jm}, CMB observations~\cite{Clark:2016nst}, and high-energy backgrounds (e.g., EGRB~\cite{Carr:2009jm} and Voyager~\cite{Boudaud:2018hqb}), excluding PBHs in the mass range $10^{9}-10^{17}\,\mathrm{g}$. At larger masses, microlensing surveys such as MACHO~\cite{Macho:2000nvd}, EROS~\cite{EROS-2:2006ryy}, Kepler~\cite{Griest:2013aaa}, OGLE~\cite{Mroz:2024mse,Mroz:2024wia,Mroz:2025xbl}, and Subaru/HSC~\cite{Smyth:2019whb} constrain PBHs through the nonobservation of transient magnification events of background sources. Together, these constraints, derived under the assumption of monochromatic PBH mass functions, leave only a narrow asteroid-mass window, as shown on the right panel of Fig.~\ref{fig:GW_and_fPBH}, where the colored regions are shown for reference only. In the following, we normalize the PBH abundance to $f_{\rm PBH}^{\rm tot}=1$ to illustrate the maximal allowed signal.~\footnote{For extended PBH mass functions, constraints are typically implemented via the integral criterion 
$I_i = \int \mathrm{d}\ln M\, f_{\rm PBH}(M)/f^{\rm mono}_{i,\max}(M) \lesssim 1$, 
where $i$ labels different observational constraints, see e.g. Carr et al.~\cite{Carr:2017jsz}.}

Substituting Eq.~(\ref{eq:horizon_mass}) into Eq.~(\ref{eq:freq_k_relation}), one finds that PBH formation in this mass range is accompanied by the generation of SIGWs with peak frequencies $f_{\mathrm{peak}}$ spanning the range $\sim (10^{-4}-10^{-1})$ Hz. These signals fall squarely within the optimal sensitivity windows of future space-based laser interferometers, such as LISA~\cite{LISA:2017pwj,Babak:2021mhe}, Taiji~\cite{Luo:2019zal}, and TianQin~\cite{Luo:2025sos,Luo:2025ewp} in the millihertz (mHz) band, as well as DECIGO~\cite{Kawamura:2006up,Kawamura:2011zz} and BBO~\cite{Crowder:2005nr,Corbin:2005ny} in the decihertz band.
We fix the characteristic scale of the BPL power spectrum to $k_\ast = 1.56\times 10^{13}\,\mathrm{Mpc}^{-1}$ with $\beta = 3$ and $s = 1$, corresponding to $M_{\rm peak} \sim 10^{-13}\,M_\odot$, and adopt a smoothing scale $r = 2.74/k_\ast$ and a collapse threshold $\mathcal{C}_c = 0.587$.

For PBHs to constitute all of the DM, the required effective amplitude is $A_{\rm eff} = A\mathcal{A}_{\phi} \simeq 0.02$ for the positive non-Gaussian case and $A_{\rm eff} \simeq -0.08$ for the negative case. Since the energy density of the induced GWs scales as $h^2\Omega_{\rm GW} \sim 10^{-5} A_{\rm eff}^4$, the corresponding stochastic backgrounds peak at amplitudes of $\sim 10^{-12}$ and $\sim 10^{-10}$, respectively, as illustrated by the dashed and dash-dotted curves in the left panel of Fig.~\ref{fig:GW_and_fPBH}.
This highlights an important observational opportunity: while asteroid-mass PBHs are challenging to detect directly, their associated GW signals lie well within the sensitivity reach of future space-based interferometers. Moreover, Fig.~\ref{fig:GW_and_fPBH} demonstrates the correlated dependence of the SIGW amplitude and PBH abundance on the effective parameter $A_{\rm eff}$. Increasing $A_{\rm eff}$ (equivalently through the non-Gaussian coefficient $A$ or the peak amplitude of the field power spectrum $\mathcal{A}_\phi$) enhances both the GW signal and the PBH abundance, whereas smaller values suppress both.

\section{Thermal Inflation as a Benchmark Realization}
\label{sec:thermal_inflation}

In this section, we present thermal inflation as a benchmark realization of the purely quadratic non-Gaussian perturbations considered in this work, embedded in a concrete early-Universe setting. The thermal inflation setup itself follows standard treatments in the literature~\cite{Dimopoulos:2019wew}. 
This scenario provides a well-motivated mechanism for generating enhanced curvature perturbations on small scales, which can give rise to both PBHs and an associated stochastic background of SIGWs. We then compute the resulting signals and confront them with current observational constraints.

Thermal inflation is driven by a scalar field, often referred to as the flaton, whose potential is approximately flat and lifted by symmetry-breaking effects that generate a soft mass term $m$.
At finite temperature, the flaton is stabilized at the origin by thermal corrections, generating an effective potential of the form
\begin{equation}
V(\phi,T) = V_0 - \frac{1}{2}(m^2 - g^2 T^2)\phi^2 + \sum_{n=1}^\infty\lambda_n\frac{\phi^{2n+4}}{M_{\rm pl}^{2n}} \,,
\label{eq:flaton_full_potential}
\end{equation}
where $g,\, \lambda_n\lesssim 1$ are perturbative coupling constants.
As the temperature decreases, the vacuum energy $V_0$ eventually dominates the energy density, and the universe undergoes a short period of secondary inflation. When the temperature drops below the critical value $T_c = m/g$, thermal corrections can no longer stabilize the origin, and the effective mass squared becomes negative. A tachyonic instability then develops, driving the flaton toward its true minimum and thereby terminating thermal inflation.
The duration of the thermal inflation epoch is quantified by the number of $e$-folds, given by
\begin{equation}
    N_T = \ln\left(\frac{T_v}{T_c}\right) = \ln\left(\frac{V_0^{1/4}}{m}\right) + \frac{1}{4}\ln\left(\frac{30}{\pi^2 g_\ast}\right) + \ln g\,,
\end{equation}
where $T_v= \left(\frac{30}{\pi^2 g_\ast}\right)^{1/4}V_0^{1/4}$ marks the temperature at which vacuum domination begins.

\subsection{Curvature power spectrum from tachyonic fluctuations}
The curvature power spectrum during thermal inflation is determined by the dynamics of flaton fluctuations near the end of the inflationary phase. Near the origin, nonrenormalizable terms in the effective potential are negligible, and the flaton potential can be well approximated by its quadratic form,
\begin{equation}
    V(\phi) \simeq V_0 - \frac{1}{2}(m^2-g^2T^2)\phi^2\,.
    \label{eq:flaton_potential}
\end{equation}
Since the temperature redshifts as $T \propto a^{-1}$ during the vacuum-dominated stage, it is convenient to normalize $a = 1$ at the critical temperature $T_c$. With this choice, the temperature-dependent effective mass squared reads 
\begin{equation}
m_{\rm eff}^2 \equiv - m^2 \left(1 - a^{-2}\right)\,.
\label{eq:m_eff}
\end{equation}
Immediately below the critical temperature ($a>1$), the negative effective mass squared initiates a tachyonic instability that drives the exponential growth of sufficiently long-wavelength flaton modes.

Following the standard canonical quantization procedure in an expanding background, we decompose the flaton fluctuation into Fourier modes and introduce the canonically normalized Mukhanov-Sasaki variable, $u_k(\eta) \equiv a(\eta) \delta\phi_k(\eta)$, where $\eta \simeq -1/(aH)$ is the conformal time. Substituting the effective mass (\ref{eq:m_eff}) into the Klein-Gordon equation and using the de Sitter background relation $a''/a = 2/\eta^2$, the equation of motion for the mode function simplifies to
\begin{equation}
    u_k''(\eta) + \left( k^2 + m^2 - \frac{\nu^2 - 1/4}{\eta^2} \right) u_k(\eta) = 0 \,, \qquad \nu^2 \equiv \frac{9}{4} + \frac{m^2}{H^2} \,,
\end{equation}
where the prime denotes the derivative with respect to $\eta$.

Defining an effective wave number $\kappa \equiv \sqrt{k^2 + m^2}$, the solution satisfying the standard Bunch-Davies vacuum condition deep inside the horizon ($-\kappa\eta \gg 1$) is given by
\begin{equation}
    u_k(\eta) = \frac{\sqrt{\pi}}{2} e^{i(\nu+1/2)\frac{\pi}{2}} \sqrt{-\eta} H_\nu^{(1)}(-\kappa\eta) \,,
\end{equation}
where $H_\nu^{(1)}$ is the Hankel function of the first kind.
As the tachyonic instability develops, the cosmologically relevant scales are stretched far outside the horizon ($- \kappa\eta \to 0$). In this superhorizon limit, the mode function is dominated by the small-argument behavior of the Hankel function, $H_\nu^{(1)}(z) \simeq -\frac{i}{\pi} \Gamma(\nu) (z/2)^{-\nu}$,
yielding the resulting power spectrum of the flaton field, $\mathcal{P}_{\delta\phi}(k) \equiv \frac{k^3}{2\pi^2} |\delta\phi_k|^2$, as
\begin{equation}
\mathcal{P}_{\delta\phi}(k) \simeq \frac{2^{2\nu-1}\Gamma(\nu)^2}{\pi}\left(\frac{H}{2\pi}\right)^2\left(\frac{k}{aH}\right)^3 \left(\frac{k^2+m^2}{a^2H^2}\right)^{-\nu}\,.
\label{eq:flaton_power_spectrum}
\end{equation}
This spectrum can be accurately identified with the BPL profile defined in Eq.~(\ref{eq:BPL_spectrum}), with the UV slope $\beta = 2\nu - 3$ and smoothness parameter $s = \nu$. In the infrared regime, $k \ll m$, the spectrum scales as $\mathcal{P}_{\delta\phi} \propto k^3$, while in the UV regime, $k \gg m$, it exhibits the asymptotic behavior $\mathcal{P}_{\delta\phi} \propto k^{3 - 2\nu}$. The peak comoving wave number $k_\ast$, which maximizes the power spectrum, is located at
\begin{equation}
k_\ast = H \sqrt{\frac{3}{2}\left(\nu + \frac{3}{2}\right)} \,.
\label{eq:peak_wave_number}
\end{equation}
The corresponding variance of the flaton field is evaluated by integrating $\mathcal{P}_{\delta\phi}(k)$ over the superhorizon modes:
\begin{equation}
\langle \delta\phi^2(\mathbf{x}) \rangle = \int_{0}^{aH} \frac{\mathrm{d}k}{k}\, \mathcal{P}_{\delta\phi}(k) \xrightarrow{t\rightarrow\infty} \frac{\Gamma \left( \nu -\frac{3}{2} \right) \Gamma \left( \nu \right)}{\sqrt{\pi}}\left( \frac{H}{2\pi} \right) ^2\left( \frac{2aH}{m} \right) ^{2\nu -3}.
\end{equation}

To compute the curvature power spectrum, we must map the flaton fluctuations to $\zeta$. Since the potential (\ref{eq:flaton_potential}) preserves $\phi \rightarrow -\phi$ symmetry, the linear contribution to the primordial curvature perturbation is absent. The leading-order contribution therefore arises entirely from quadratic fluctuations, naturally yielding a non-Gaussian curvature perturbation.
In the spatially flat gauge, the curvature perturbation $\zeta$ on uniform density hypersurfaces is given by
\begin{equation}
\zeta = - H \frac{\delta\rho}{\langle\dot{\rho}\rangle} \simeq - \frac{H m_{\rm eff}^2}{\partial_t \left[m_{\rm eff}^2\langle\phi^2\rangle\right]}\delta\phi^2\,,
\label{eq:zeta}
\end{equation}
where $\delta\phi^2 \equiv \phi^2 - \langle\phi^2\rangle$.
Assuming that $\delta\phi$ is a Gaussian random field, the power spectrum of $\delta\phi^2$ can be computed using Wick's theorem as
\begin{equation}
\mathcal{P}_{\delta\phi^2}(k)
= \frac{k^3}{2\pi} \int_0^{aH} \mathrm{d}^3q\, \frac{\mathcal{P}_{\delta\phi}(q)\mathcal{P}_{\delta\phi}(|\mathbf{k}-\mathbf{q}|)}{q^3|\mathbf{k}-\mathbf{q}|^3}
= \int_{0}^{aH/k} \mathrm{d}x \int_{|1-x|}^{1+x} \mathrm{d}y \, \frac{1}{x^2 y^2} \mathcal{P}_{\delta\phi}(kx) \mathcal{P}_{\delta\phi}(ky) \,,
\label{eq:quadratic_power_spectrum}
\end{equation}
where we have introduced the dimensionless variables $x \equiv q/k$ and $y \equiv |\mathbf{k}-\mathbf{q}|/k$. 
By substituting Eq.~(\ref{eq:flaton_power_spectrum}) into Eq.~(\ref{eq:quadratic_power_spectrum}), one finds that the explicit time-dependent terms drop out in the late-time limit, leading to a constant curvature power spectrum:
\begin{equation}
\mathcal{P}_\zeta(k)\equiv \frac{k^3}{2\pi} \int\mathrm{d}^3x\, e^{-i\mathbf{k}\cdot\mathbf{x}}\langle\zeta(0)\zeta(\mathbf{x})\rangle \xrightarrow{t\to\infty} \frac{2}{\pi(\nu-1)} \left[ \frac{\Gamma(\nu)}{\Gamma(\nu-1/2)} \right]^2 \alpha^{4\nu-6} \int_0^{\infty} \mathrm{d}z \, \frac{z}{(z^2+\alpha^2)^{\nu}} \Big[ \mathcal{F}(1-z) - \mathcal{F}(1+z) \Big] \,,
\end{equation}
where we have defined
\begin{equation}
\alpha \equiv \frac{m}{k} \quad \text{and} \quad \mathcal{F}(w) \equiv \left( w^2 + \alpha^2 \right)^{1-\nu}.
\end{equation}
A straightforward asymptotic analysis of the above expression reveals that the curvature power spectrum retains a BPL profile, characterized by a steep $k^3$ ascent on large scales and a $k^{3-2\nu}$ fall-off on small scales.

After thermal inflation, the universe undergoes a fast-roll inflationary phase that generates additional expansion before reheating.\footnote{For $m \sim H$, the field evolves in a fast-roll regime with $\dot\phi \sim \mathcal{O}(H)\phi$. Although the slow-roll condition is violated, the energy density can still be temporarily dominated by the vacuum energy near the origin. In this regime, $\epsilon_H = \dot\phi^2 / (2M_{\rm pl}^2 H^2) \sim (\phi / M_{\rm pl})^2 \ll 1$ so accelerated expansion persists before the kinetic energy becomes important.} The total number of $e$-folds accumulated during this stage, denoted by $N_F$, can be estimated as~\cite{Linde:2001ae}
\begin{equation}
    N_F \simeq \ln\left(\frac{m}{H}\right) + \frac{1}{2\nu-3}\ln\left[\frac{3\times 4^{3-\nu}\pi^{5/2}M_{\rm pl}^2}{\Gamma(\nu-3/2)\Gamma(\nu)m^2}\right].
\end{equation}
The reheating temperature is obtained by assuming that the vacuum energy stored in the flaton field is instantaneously converted into radiation once the decay width $\Gamma_{\rm dec}$ becomes comparable to the Hubble parameter:
\begin{equation}
T_{\rm reh} = T_v \sqrt{\frac{\Gamma_{\rm dec}}{H}} = \xi^{1/2} T_v \,, \qquad \xi \equiv \Gamma_{\rm dec}/H\,.
\end{equation}
Here, $H \simeq \sqrt{V_0/(3M_{\rm pl}^2)}$ represents the Hubble scale driven by the flaton vacuum energy.
For the sake of comparison with the conventional normalization $a_0=1$, we evaluate the present-day scale factor $a_0$ in our normalization, as determined by the subsequent expansion history:
\begin{equation}
    a_0 = e^{N_F}\xi^{-2/3} \left[\frac{s(T_{\rm reh})}{s(T_0)}\right]^{1/3} 
    \simeq 7.40 \times 10^{12} e^{N_F}\xi^{-1/6} \frac{g_{\ast,r}^{1/12}}{g_{\ast,0}^{1/3}} \left( \frac{\sqrt{H M_{\rm pl}}}{1\,\text{GeV}} \right).
\end{equation}

\subsection{Phenomenological implications: PBHs and SIGWs}

We now turn to the phenomenological implications of the curvature perturbations generated during thermal inflation, focusing on the resulting PBH abundance and the associated SIGW spectrum. To this end, we consider two representative parameter choices corresponding to characteristic scales relevant for PTA observations and for the sensitivity range of space-based interferometers. We evaluate the PBH mass fraction within the extended Press-Schechter formalism by setting the smoothing scale to $r = 2.74 / k_\ast$ throughout.

\begin{figure}[htbp!]
    \centering
    \includegraphics[width=.48\columnwidth]{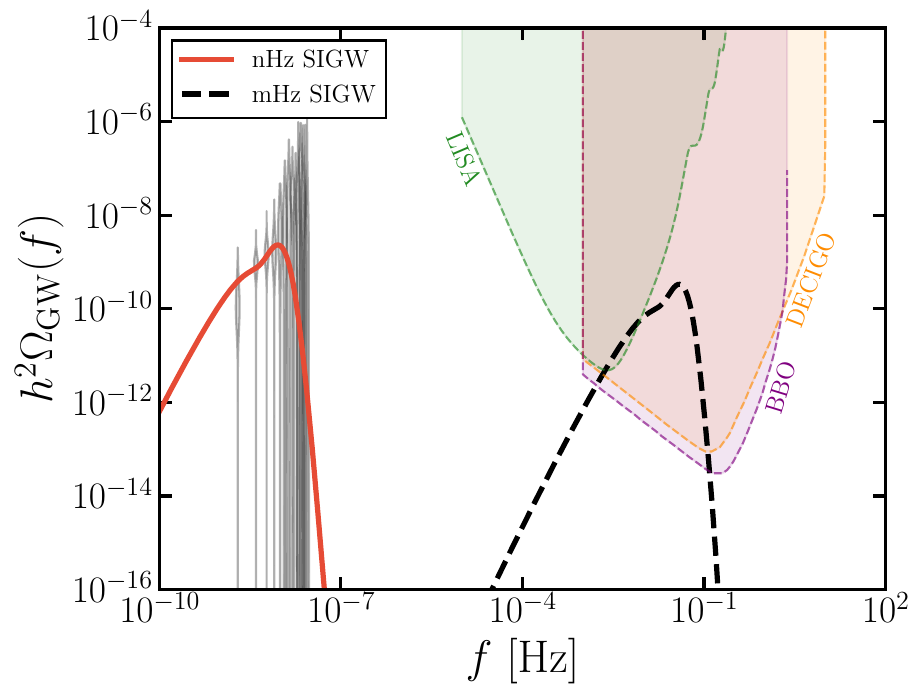}%
    \hfill
    \includegraphics[width=.48\columnwidth]{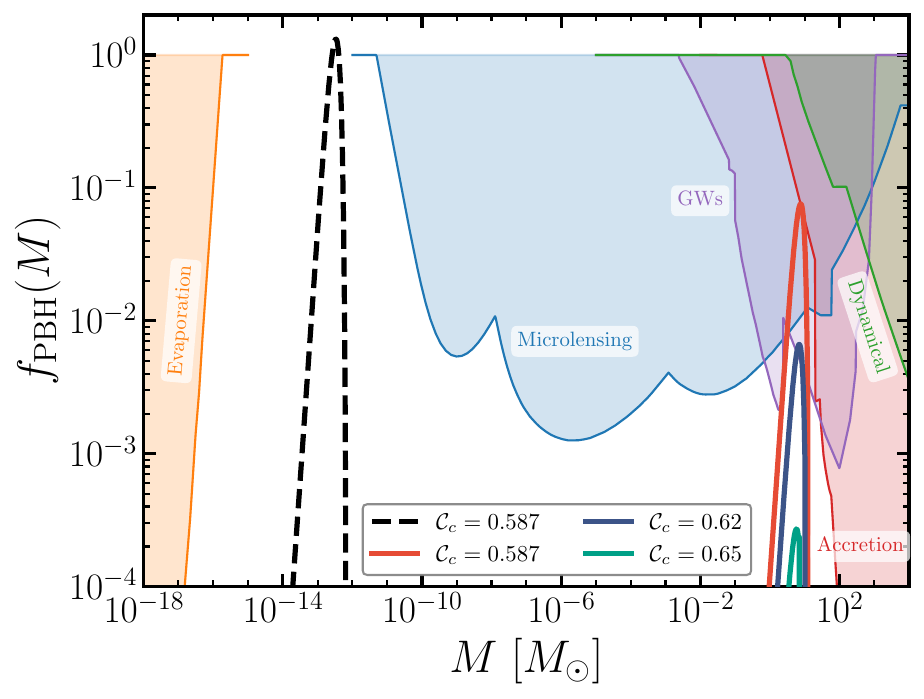}
    \caption{\label{fig:SIGW-PBH}%
    \textit{Left panel:} SIGW spectra $h^2\Omega_{\rm GW}(f)$ as a function of frequency $f$. The red solid curve shows a PTA-compatible signal fitted to the NANOGrav 15-year data in the nHz band, while the black dashed curve illustrates a spectrum peaking in the mHz band, relevant for space-based interferometers. Shaded regions indicate the projected sensitivities of LISA (green), BBO (purple), and DECIGO (orange).
    \textit{Right panel:} Extended PBH mass distributions $f_{\rm PBH}(M)$ compared with observational constraints. The black dashed curve shows a spectrum peaking in the asteroid-mass window for $\mathcal{C}_c = 0.587$. The colored solid curves show stellar-mass spectra for $\mathcal{C}_c = 0.587$ (red), $0.62$ (blue), and $0.65$ (green), illustrating the sensitivity of the PBH abundance to the collapse threshold.}
\end{figure}

As a representative setup for PBH DM, we fix parameters to $m = 10^3\,\text{GeV}$, $m/H = 9.25$, and $\xi = 10^{-14}$. The resulting SIGW spectrum, represented by the black dashed curve in the left panel of Fig.~\ref{fig:SIGW-PBH}, peaks prominently in the mHz to decihertz band. This presents a highly promising observational target, as the signal falls perfectly within the sensitivity range of upcoming space-based interferometers such as LISA, DECIGO, and BBO. The corresponding PBH mass distribution in the right panel, computed with a fiducial compaction threshold $\mathcal{C}_c=0.587$, is entirely localized within the unconstrained asteroid-mass window. Here, the parameters naturally allow the PBHs to account for the total DM abundance ($f_{\rm PBH}^{\rm tot} \simeq 1$) while safely evading all current astrophysical constraints, including Hawking evaporation and microlensing limits.
This demonstrates that purely quadratic NG generated via the tachyonic instability at the end of thermal inflation does not generically suppress PBH formation, but can instead support efficient collapse depending on the spectral shape.

In contrast, the second benchmark setup with $m = 1.67\times 10^{-8}\,\text{GeV}$, $m/H = 6.11$,  and $\xi = 1$ yields a GW spectrum that peaks in the nHz band with an amplitude compatible with current PTA sensitivities. However, for this parameter choice, satisfying the PBH abundance constraint comes at the expense of a good fit to the NANOGrav 15-year dataset. This tension is indicative of a more general feature of thermal inflation scenarios. It is challenging to identify a parameter region that simultaneously reproduces the PTA signal and avoids PBH overproduction. This tension persists even under variations of the compaction threshold $\mathcal{C}_c$, as illustrated in Fig.~\ref{fig:SIGW-PBH}.

This limitation can be traced to the fact that the thermal inflation model effectively contains only a single controlling parameter, $m/H$, which simultaneously determines the UV slope $\beta = 2\nu - 3$ and the smoothness parameter $s = \nu$, as follows from Eqs.~(\ref{eq:flaton_power_spectrum}) and (\ref{eq:peak_wave_number}). To account for the PTA signal, $A_{\rm eff}$ must lie within the PTA-preferred range. Combined with the relation $A_{\rm eff} \simeq -1/(2\nu - 3)$, this fixes $\nu$ and hence leads to relatively small $\beta$ and $s$, corresponding to a broad power spectrum.
As shown in our analysis of purely quadratic non-Gaussian perturbations, such a spectrum is insufficient to drive $\rho$ close to $-1$, and therefore fails to suppress the exponential tail of the compaction-function distribution. Consequently, PBH formation remains too efficient within this thermal inflation framework, preventing a successful resolution of the PTA–PBH tension.
This issue may be alleviated if the tachyonic phase is made more transient. For example, incorporating backreaction effects that terminate the instability at an early stage can effectively truncate the growth of modes, leading to a sharper spectrum with an exponential cutoff.

\section{Conclusions and Discussion}
\label{sec:summary}

In this paper, we have investigated PBH formation and SIGWs in scenarios where curvature perturbations are generated by tachyonic amplification with purely quadratic NG. Within this framework, we derived the probability distribution of the compaction function and obtained analytic expressions for the PBH mass spectrum and abundance, highlighting the crucial role of non-Gaussian statistics in shaping the tail of large perturbations.

A central result of this work is that the probability distribution of the compaction function exhibits an exponential tail, such that the PBH abundance becomes exponentially sensitive not only to the variance of perturbations, but also to their correlation structure. In particular, we find that the suppression of the tail is controlled by the correlation coefficient $\rho$, revealing a qualitatively new feature of purely quadratic non-Gaussian scenarios in which correlation effects play a central role beyond conventional Gaussian estimates.

We further show that the spectral width of the curvature power spectrum plays a decisive role in determining the correlation structure and the efficiency of PBH formation. Broad spectra, characterized by moderately negative correlation coefficients, leave the exponential tail unsuppressed, thereby enhancing PBH formation and allowing PBHs to account for DM in suitable parameter regions. In contrast, sufficiently narrow spectra naturally drive $\rho \to -1$ and strongly suppress PBH formation, providing a general mechanism for alleviating the PTA–PBH tension.

A concrete realization of this framework is provided by thermal inflation. In this setup, spectra compatible with PTA observations are typically broad, which makes it difficult to simultaneously reproduce the PTA signal while avoiding PBH overproduction. This illustrates a limitation of thermal inflation in addressing the PTA–PBH tension. An abrupt termination of tachyonic growth, for example through additional interactions or backreaction effects, can in principle generate a UV cutoff and sharply peaked spectra, offering a viable mechanism to suppress PBH formation. We leave a detailed investigation of such scenarios for future work.

We also identify two distinct classes of tachyonic amplification, corresponding to monotonic and nonmonotonic growth histories, which produce BPL and LN spectra, respectively. This classification clarifies how the growth history of the tachyonic instability determines the spectral shape and, consequently, the efficiency of PBH suppression.

A key theoretical uncertainty lies in the collapse threshold and smoothing scale entering the compaction formalism, which are not well determined for strongly non-Gaussian perturbations and may depend sensitively on the shape of the overdensity profile. This highlights the need for dedicated numerical simulations to calibrate PBH formation in such scenarios.

Another theoretical issue concerns the validity of the purely quadratic approximation. The purely quadratic relation used in this work represents the leading nonvanishing contribution generated by the tachyonic instability mechanism. Higher even-order terms may arise once the full nonlinear relation between the tachyonically amplified field and the curvature perturbation is taken into account. In the thermal inflation realization, however, the relevant amplification occurs while the flaton remains close to the origin, where the effective potential in Eq.~(\ref{eq:flaton_full_potential}) is well approximated by the quadratic term. The stabilizing nonlinear terms become important only near the end of the fast-roll stage, so the induced higher even-order contributions to the curvature perturbation are expected to be subleading during the main perturbation-generation epoch. Since PBH formation is sensitive to the extreme tail of the distribution~\cite{Young:2013oia,Ferrante:2022mui}, these terms may nevertheless modify the quantitative abundance if the tail probes the nonlinear regime. A systematic study of these model-dependent corrections remains an important avenue for future investigations.

Overall, our results demonstrate that purely quadratic NG provides a useful framework for linking tachyonic instability, PBH formation, and SIGWs. The exponential sensitivity of the PBH abundance to the correlation coefficient $\rho$ makes the spectral shape, rather than only the amplitude, a crucial ingredient in resolving the PTA--PBH tension. These findings point to a promising direction for connecting early-Universe physics with future observations of GWs and PBH dark matter.

\section*{Acknowledgments} 

We are grateful to Hong-An Zeng and Xuan Liu for helpful discussions. This work was supported in part by the National Natural Science Foundation of China (NSFC) under Grants No.~12235016 and No.~12221005, and by the Fundamental Research Funds for the Central Universities.

\section*{Data availability}

There are no publicly available research data or software supporting this manuscript. Requests for further information or data should be sent to the authors.

\bibliography{refs}

\end{document}